\documentclass[a4paper,11pt]{article}
\usepackage{jheppub}
\usepackage{latexsym,amsmath,amsfonts,amssymb}
\usepackage{epsfig,graphics,graphicx,pgf,pstricks}
\usepackage{animate}
\usepackage{comment}
\usepackage{color}
\usepackage{ulem}

\newcommand{\be}{\begin{equation}}
\newcommand{\ee}{\end{equation}}

\newcommand{\tQ}{{\tilde Q}}

\newcommand{\cO}{{\cal O}}

\newcommand{\cH} {{\cal H}}

\newcommand{\cP} {{\cal P}}

\newcommand{\vev}[1] {\left<#1\right>}

\title{On Democratic String Field Theories}
\author{Stefano Giaccari$^{(1)}$, Michael Kroyter$^{(2)}$}
\affiliation{$^{(1)}$Dipartimento di Fisica e Astronomia ‘Galileo Galilei’ e INFN sez. di Padova,\\
\hphantom{$^{(1)}$}Via Marzolo 8, 35131 Padova, Italy\\
$^{(2)}$Department of Mathematics, Holon Institute of Technology,\\
\hphantom{$^{(2)}$}Golomb St. 52, Holon 5810201, Israel}
\emailAdd{stefano.giaccari@pd.infn.it}
\emailAdd{michaelkro@hit.ac.il}

\abstract{We reexamine democratic open string field theories,
					namely, theories in which string fields are not constrained to a single picture number and picture changing is
					obtained as a gauge transformation. We describe several possibilities for regular free theories
					and attempt to construct the lowest order interaction term and identify the lowest order gauge transformation
					for some of these theories. We also discuss projections over string field spaces that might be needed for
					a consistent off-shell implementation of picture changing.}

\keywords{String Field Theory, Superstrings and Heterotic Strings}

\begin{document}

%=========
\maketitle
%=========

%=====================
\section{Introduction}
%=====================
\label{sec:intro}

Since the construction of a covariant open string field theory by Witten~\cite{Witten:1985cc}, attempts have been made to find
a similar framework for the superstring. Witten's own proposal~\cite{Witten:1986qs}, which makes use of the BRST  operator $Q$,
together with the picture changing operator (PCO) $X$ and the inverse PCO, $Y$ of the Ramon-Neveu Schwarz (RNS) formalism,
turned out to be inconsistent~\cite{Wendt:1987zh} and modifications of this proposal~\cite{Preitschopf:1989fc,Arefeva:1989cp} 
were shown to suffer from similar problems~\cite{Kroyter:2009zi}.
The inconsistencies manifest themselves as singularities in the form of the gauge transformations and the propagators of these theories,
where string fields are multiplied using Witten's half-string product and $X$ is inserted at the vertex midpoint.
The source of the problems is the divergences due to the collision of PCOs inserted at the mid-point, that is, it is related to the
special role played by the mid-point in theories based on Witten's star product and to the existence of picture number
in the RNS formulation of the superstring~\cite{Friedan:1985ge}. Explicitly, since the picture number is related to a
redundancy in the description of vertex operators, the common lore was that a string field should be fixed to a particular value of
the picture number.
Then, PCOs had to be added to the action and to the gauge transformation.
In order to retain the algebraic structure of the bosonic theory the PCOs had to be inserted in the string mid-point.
Whereas the mentioned problems were resolved for the Neveu-Schwarz (NS) sector in Berkovits' formulation~\cite{Berkovits:1995ab},
a covariant solution for the Ramond sector was still lacking due to the different picture numbers that are
assigned to the different sectors.

A different approach was introduced in~\cite{Kroyter:2009rn} (see also~\cite{Kroyter:2010rk,Kroyter:2011nc}).
Here, the picture number is not fixed and the redundancy is treated as a gauge symmetry.
This approach was named ``democratic'', since all pictures can influence the physics.
The democratic theory resolved the problems with the gauge transformations while naturally unifying the NS and Ramond (R) sectors.
Other formulations of superstring field theory can be obtained as partially gauge fixed versions of the democratic theory.
Hence, one could claim that beside the natural unification of the various sectors this theory has the additional advantage of
manifesting a symmetry of the first quantized theory.

Another approach was later introduced by Erler, Konopka, and Sachs (EKS)~\cite{Erler:2013xta}.
There, the mid-point insertion was replaced by a smeared form of the PCO, which is naturally obtained from
the line integral of the $\xi$ operator that was introduced in~\cite{Iimori:2013kha} for relating the Berkovits formulation in the large
Hilbert space and a regular formulation in the small Hilbert space.
This construction resolved the singularities. The price to be paid was that
the theory was no longer cubic. Instead an $A_\infty$ structure had to be introduced.
Following~\cite{Erler:2013xta}, many papers appeared with improved proposals for superstring field theory actions
and relations among different theories (see, e.g.,~\cite{Erler:2014eba,Erler:2015lya} and~\cite{Erbin:2021smf,Maccaferri:2023vns,Sen:2024nfd}
for recent general reviews of string field theory).

While this construction can be generalized to include also the Ramond sector at the level of the
equations of motion (EOM)~\cite{Erler:2015lya},
it is not so simple to define a kinetic term at the level of the action for the Ramond sector, since it is related to the notoriously
hard problem of integration over the zero modes,
which can be taken into account by restricting the state space appropriately~\cite{Kazama:1985hd,Kugo:1988mf}.
A fundamental breakthrough came from the realization that interactions may be consistent with this projection.
This allowed Sen to construct the 1PI effective superstring field theory including the Ramond sector in~\cite{Sen:2015hha}.
As a further development, complete gauge invariant theories have been constructed, either bases on Berkovits'
large Hilbert space theory~\cite{Kunitomo:2015usa}
or on the small Hilbert space approach of EKS, exhibiting the $A_\infty$ structure~\cite{Erler:2016ybs,Konopka:2016grr}.

It should be noted, that in all these significant advances, the democratic approach did not play any role,
presumably since it was considered to be singular in some sense.
In this work we would like to reexamine the democratic theory and suggest new variants thereof.
A consistent regular democratic theory, if found, could in principle provide us, in a regular setting,
with the benefits of the original democratic
theory, and more. Namely, it could unify more naturally different sectors of the theory, identify some hidden novel
mathematical symmetries or structures of superstring theory. Partial gauge fixings of such a theory might reduce it to the various
new known string field theory formulations, as well as to other novel string field theory formulations, that could be useful in
various contexts. It might also be useful for the construction of a master action.
Then, partial gauge fixings might be used in order to identify the master action of other known theories, e.g., to improve the results
of~\cite{Kroyter:2012ni,Berkovits:2012np,Berkovits:XXX,Iimori:2015aea}.

Naturally, such a construction is all but simple and this work is merely a first step in this direction.
As in the construction of other string field theories we consider a perturbative approach
and attempt to identify only the lowest order interaction term in the NS sector.

The rest of the paper is organized as follows:
In section~\ref{sec:old} we review possible worldsheet formulations of the cohomology problem
of the open superstring that could be used as a starting point towards the construction of superstring field theory,
as well as the formulation of~\cite{Kroyter:2009rn} (the ``old democratic theory'') and some of its problems.
Next, in section~\ref{sec:preliminaries}, we describe some preliminaries needed for the construction, in particular,
PCOs and a projection over the space of string fields. Then, in section~\ref{sec:XY} we examine the properties of
the operator defining the proposed projection. A free action is constructed in section~\ref{sec:free} and is shown
to produce the expected results. Two different approaches for the construction of the lowest order interaction term
are presented in sections~\ref{sec:cubic} and~\ref{sec:partialFixed} respectively,
where we also attempt to determine the lowest order non-linear term of the gauge transformation and identify a subtlety
with its definition.
A different approach towards defining a theory is briefly described in section~\ref{sec:peoples} and finally, some concluding remarks
are offered in section~\ref{sec:conclusions}.

%=============================================================
\section{The cohomology problem and the old democratic theory}
%=============================================================
\label{sec:old}

In this section we present some relevant background material. In~\ref{sec:cohomology} we introduce several options
for defining the cohomology problem of the open superstring, with the understanding that specifying a cohomology problem
is the first step towards the formulation of a universal string field theory. We also describe various PCOs.
Next, in~\ref{sec:oldDemocratic} we describe the construction of the democratic theory. Then, in~\ref{sec:crit}
we present some criticism on the theory which we attempt to address in later sections.

%===========================================
\subsection{The cohomology problem and PCOs}
%===========================================
\label{sec:cohomology}

For constructing a string field theory one has to define a proper off-shell generalization of vertex operators.
To that end, the cohomology problem defining the vertex operators of the theory should be defined.
However, different equivalent representations of the cohomology problem exist and different string field
spaces that generalize the various options can be defined.
In what follows we consider the NS sector of the open superstring in the bosonized variables~\cite{Friedan:1985ge}.
When using these variables one must distinguish between the small Hilbert space $\cH_S$ and the large Hilbert space $\cH_L$.
As usual, the small Hilbert space is defined by the constraint $\eta V=0$, where, $\eta\equiv \eta_0$ is the zero mode of the $\eta$ ghost.
This space can be decomposed as a direct sum of spaces with fixed ghost and picture numbers,
\begin{equation}
\cH_S=\bigoplus_{g,p} \cH_S^{g,p}\,.
\end{equation}

In its most familiar definition, the cohomology problem of the open NS string is given by considering the space
of ghost number $g=1$ states with an arbitrary fixed integer picture number $p$ in the small Hilbert space.
The cohomology problem is then defined by,
\begin{equation}
\label{standardCoho}
QV=0\,,\qquad V\sim V+Q\Lambda\,,
\end{equation}
with $V\in \cH_S^{1,p}$ and $\Lambda \in \cH_S^{0,p}$ for arbitrary fixed integer picture number $p$.
This is the starting point (with particular choices of $p$) for several string field theory formulations,
e.g.~\cite{Witten:1986qs,Preitschopf:1989fc,Arefeva:1989cp,Erler:2013xta}.

The EOM $QV=0$ and the requirement of being in the small Hilbert space $\eta V=0$ can
switch their role. Instead of thinking of the latter as defining what we call the small Hilbert space and defining
the cohomology problem of $Q$ in this space, we can use the former to define the dual small Hilbert space~\cite{Kroyter:2009rn}
and study the cohomology problem of $\eta$ in this space, with a proper fixing of picture number.
This is trivially equivalent to the standard cohomology problem and is explicitly given by,
\begin{equation}
\label{dualCoho}
\eta V=0\,,\qquad V\sim V+\eta\hat\Lambda\,,
\end{equation}
with $V$ and $\hat \Lambda$ in the dual small Hilbert space, defined explicitly by constraining these large Hilbert space
operators by,
\begin{equation}
Q V=0\,,\qquad Q\hat\Lambda=0\,.
\end{equation}

String field theories based on this formulation are easily constructed by extending the duality between $Q$ and $\eta$
to the non-linear level~\cite{Kroyter:2010rk}.
A salient point of both these definitions of the cohomology problem is the fact that the gauge parameters are constrained.
In $\cH_L$ it becomes, however, apparent that the two cohomological problems are actually the same
\begin{equation}
\label{Coho1}
Q V=\eta V=0\,,\qquad
V\sim V+Q\eta \tilde \Lambda\,,
\end{equation}
where $\tilde \Lambda\in \cH_L$ is only constrained to have given ghost and picture numbers.
Another observation is the fact that both $Q$ and $\eta$ have trivial cohomology in $\cH_L$, so that given
a vertex\footnote{Where it is clear from the context what the ghost number is, as is the case here, we identify string field spaces
and vertices/string fields living in these spaces solely by their picture number.} $V_{-1}$, one can define
\begin{equation}
\label{V0Vm2}
V_{0}=Q\left(\xi V_{-1}\right)\,,\qquad
V_{-2}=\eta\left(P V_{-1}\right)\,,
\end{equation}
where we defined,
\begin{equation}
P\equiv -c\xi\partial\xi e^{-2\phi}\,.
\end{equation}
Note, that we are using the standard RNS bosonized variables, the fields are point insertions at
arbitrary points on the worldsheet and normal ordering is understood.
$V_{0}$ and $ V_{-2}$ are still solutions of the cohomological problem described above and can themselves be used to
iterate the procedure so that one gets an infinite number of vertices
$V_p$, all corresponding to the same physical state.

The fact that the same cohomology is obtained for arbitrary integer picture number implies that homomorphisms exist
among these cohomology spaces.
This only means that there exist operators (in the linear algebraic sense)
$X_{p,q}:\cH_S^{1,p}\rightarrow \cH_S^{1,q}$ such that $X_{p,q}$ sends $g=1$ picture $p$ states to $g=1$ picture $q$ states and
the following holds,
\begin{equation}
\begin{aligned}
\label{cohoTrans}
QV_{1,p} =0 \quad &\Rightarrow \quad QX_{p,q}V_{1,p}=0\,,\\
QV_{1,q} =0 \quad &\Rightarrow \quad \exists V_{1,p}, V_{0,q}: X_{p,q}V_{1,p}= V_{1,q}+QV_{0,q}\,,\\
X_{p,q} Q V_{0,p} = V_{1,q} \quad &\Rightarrow \quad \exists V_{0,q}: V_{1,q} = Q V_{0,q}\,,\\
X_{p,q} V_{1,p} = QV_{0,q} \quad &\Rightarrow \quad \exists V_{0,p}: V_{1,p} = Q V_{0,p}\,.
\end{aligned}
\end{equation}
Note that the $X_{p,q}$ need be neither surjective nor injective.

In fact, we know that a stronger condition than just having linear transformations $X_{p,q}$ obeying~(\ref{cohoTrans}) holds.
The $X_{p,q}$ can be realized using multiplication by conformal operators. Moreover, these operators can be chosen to
depend only on $q-p$. That is, there exist operators (in the CFT sense), $X_p$,\footnote{The $z$ dependence of $X_p(z)$ is irrelevant at the level
of on-shell vertex operators.} such that all the mappings can be realized by
$X_{p}:\cH_S^{1,q-p}\rightarrow \cH_S^{1,q}$. Furthermore, the $X_p$'s  can be chosen to be zero weight primaries~\cite{Kroyter:2009rn}.
In particular we have\footnote{The commutator in~(\ref{XQxi}) and elsewhere in this paper denotes the graded commutator,
which is an anti-commutator in~(\ref{XQxi}).},
\begin{subequations}
\label{XQxi}
\begin{eqnarray}
X_1&\equiv& X \equiv [Q,\xi]\,,\\
X_0&=&1\,,\\
X_{-1}&\equiv& Y
\equiv\left[\eta,P\right]=c \partial \xi e^{-2\phi}\,.
\end{eqnarray}
\end{subequations}
A crucial property is 
\begin{equation}
\label{XYope}
:XY:=1\,,
\end{equation}
where $::$ stands for the standard normal ordering. Note that~\eqref{XYope} does not imply that taking a normal ordered product with
$Y$ is an inverse operation to taking a normal ordered product with $X$ because of the non-associativity of the normal ordered product.
Nevertheless~\eqref{XYope} implies that $Y$ is the inverse of $X$ in the cohomology of $Q$ and/or $\eta$, i.e.,
\begin{equation}
\label{XYcoho}
\left[Y\right]\cdot\left[X\right]=1\,,
\end{equation}
where $\left[\phi\right]$ denotes the cohomology class of $\phi$ and,
given two fields ${\cal O}_1$ and ${\cal O}_2$, the dot product is defined  as
\begin{equation}
\left({\cal O}_1 \cdot {\cal O}_2\right)(w)=\frac{1}{2\pi i}
 \oint_{C_w} \frac{{\cal O}_1(z) {\cal O}_2(w)}{z-w}dz\,.
\end{equation}
Both $X$ and $Y$ have non-zero kernels, but this is not an obstruction to~\eqref{XYcoho} as long as all the fields in the kernel are
either exact or non-closed.
As CFT operators the $X_{p}$'s might have OPE singularities. This is the case already for $X$ and $Y$ that behave as,
\begin{equation}
\label{XYsing}
X(z)X(0)\sim\frac{...}{z^2}\,,\qquad Y(z)Y(0)\sim\frac{...}{z^2}\,,
\end{equation}
where the dots stand for non-trivial (but $Q$-exact) conformal operators.
Indeed, these singularities played important role in the construction
of superstring field theories, in particular as obstructions towards such constructions\footnote{These singularities play an important
role also in the definition of perturbation theory of the superstring~\cite{Witten:2012bh,Sen:2015hia}.}.
In the following we mention some approaches for avoiding these singularities.

The space of vertex operators does not possess a natural inner product.
In particular, there is no canonical way to fix the relative normalization of vertex operators describing
different physical states.
However, the existence of $X$ allows us to define a relative normalization for states describing the same physical state
at different picture numbers. While this is still not canonical, it follows from fixing the normalization of the PCO $X$.
A different choice, e.g., $\alpha X$ instead of $X$ for some $\alpha\neq 0$ would lead to different relative
normalization (and even more so in the case of choosing more general transformations $X_{p,q}$).
Here, we choose the standard form of PCOs~(\ref{XQxi}).
This fixes the form of all the $X_p$ up to (presumably singular)
$Q$-exact terms and hence the relative normalization of all the physical operators at different picture numbers.
Specifically, for any physical state $V$ we say the $X_p V$ is of the same ``norm'' as $V$.

While~(\ref{standardCoho}) is the most standard formulation of the cohomology problem, an equivalent formulation
can be obtained by considering ghost number zero states with an arbitrary fixed integer picture number
in the large Hilbert space.
Now, the cohomology problem is defined by,
\begin{equation}
Q\eta V=0\,,\qquad V\sim V+Q\Lambda+\eta\hat\Lambda\,.
\end{equation}
Again, the ghost and picture numbers of $\Lambda$ and $\hat \Lambda$ are constrained.
This formulation is the starting point for Berkovits' string field theory~\cite{Berkovits:1995ab}.

Another possibility for defining the cohomology problem uses the operator
\begin{equation}
\label{Qtilde}
\tilde Q\equiv Q-\eta\,.
\end{equation}
From
\begin{equation}
[Q,\eta]=0\,,
\end{equation}
if follows that,
\begin{equation}
\label{Qeta}
\tilde Q^2=0\,,
\end{equation}
which implies that $\tQ$ can be considered as a cohomology operator.
Since this operator includes $\eta$ we should study its cohomology in the large Hilbert space\footnote{In the small
Hilbert space this operator is identical to $Q$.}.
We further restrict the cohomology problem to ghost number one states with
an arbitrary fixed integer picture {\it or with an arbitrary finite range of integer pictures}, that is we consider,
\begin{equation}
\label{demoVertex}
\tQ V=0\,,\qquad V\sim V + \tQ\Lambda\,,
\end{equation}
with proper restrictions on $\Lambda$.
It turns out that the same cohomology problem is obtained regardless of the range of pictures taken~\cite{Berkovits:2001us}.
The point is that since $\Lambda$ is also defined in the large Hilbert space, it can be used to change the picture
number using generalizations of~(\ref{V0Vm2}).
Thus, {\it the symmetry among different picture numbers turns into a gauge symmetry}.
This is the starting point for the construction of the democratic theory.

When the restriction of a finite picture range is lifted the nature of the cohomology problem becomes more subtle
and it depends on the particular definition of the space of allowed states and allowed gauge transformations.
Without any restrictions the cohomology of the operator $\tQ$ becomes trivial~\cite{Berkovits:2001us}.
This results from the existence of contracting homotopy operators for the operator $\tQ$\footnote{A contracting homotopy
operator $A$ of a derivation $d$ obeys $dA=1$. This implies that the cohomology of $d$ is trivial, since $dV=0$ implies $V=d(AV)$.
Hence, the spaces of states and of gauge states should be appropriately restricted.
In the current case $d$ is $\tQ$ and $A$ can be $\alpha \cO_+ + (1-\alpha) \cO_- + \tQ \Lambda$ for an arbitrary constant $\alpha$
and an arbitrary ghost number $-1$ state $\Lambda$.
Similar issues appear in the study of the pure spinor formulation of string theory~\cite{Berkovits:2005bt}.},
\begin{subequations}
\label{Opm}
\begin{align}
\label{Op}
\cO_+ &\equiv -\sum_{p=1}^\infty\cO_p\,,\\
\cO_- &\equiv \sum_{p=-\infty}^0 \cO_p\,.
\end{align}
\end{subequations}
Here, the $\cO_p$ are operators that up to (singular) exact terms are given by
\begin{equation}
\cO_p\simeq \xi X_{p-1}\,.
\end{equation}
It was shown in~\cite{Kroyter:2009rn} that these operators can be chosen to be zero weight primaries obeying,
\begin{equation}
\label{OxiX}
Q\cO_p=X_p\,,\qquad \eta \cO_p=X_{p-1}\,.
\end{equation}
In particular,
\begin{equation}
\cO_0 = \xi Y = P\,,\qquad \cO_1 = \xi\,,
\end{equation}
are the contracting homotopy operators in the large Hilbert space for $Q$ and for $\eta$ respectively.

The fact that a completely unconstrained picture number leads to a trivial cohomology does not imply that
one must consider only a bounded range of picture numbers. Consider for example the space of all states with
bounded picture number. While every state has a bounded support in picture number the space itself 
is unbounded.
Disregarding any other subtlety with the definition of the space of string fields we can formally define this space
of compactly supported picture numbers by
(note that the union here is not a disjoint one),
\begin{equation}
\label{Hcom}
\cH_{com}=\bigcup_k^\infty\Big(\bigoplus_{p=-k}^k\cH_p\Big).
\end{equation}
It is clear that for $V\in\cH_{com}$, states of the form $AV$, with $A$ a contracting homotopy operator of $\tQ$,
are generically not in this space and the cohomology problem remains the standard one.
One could go further and attempt, instead of a compact support with respect to the picture number, a space including only states with
components that decay fast enough as a function of picture number. This is less trivial since, as already mentioned,
there is no natural norm in the large Hilbert space and hence it is hard to define the notion of ``decay''.
Nonetheless, in particular cases this statement can be made more
concrete, in which case the cohomology problem would again be the standard (non-trivial) one. We return to this point in what follows.

%===========================================
\subsection{Reviewing the democratic theory}
%===========================================
\label{sec:oldDemocratic}

The democratic theory of~\cite{Kroyter:2009rn} is an interacting off-shell extension of the formulation
of the cohomology problem given in~(\ref{demoVertex}).
The space of string fields is assumed to be some space that includes (off-shell) states at arbitrary picture number
subject to a constraint (that is not explicitly specified) that retains the proper cohomology problem.
The EOM is assumed to be identical in form to the bosonic one\footnote{We set for simplicity $g=1$.
It is clear how to restore it, if needed, either here or in front of the action.
String fields are multiplied with Witten's star product that we keep implicit for now. Similarly, the commutator in~(\ref{oldGaugeTrans})
is with respect to the star product.},
\begin{equation}
\label{oldEOM}
\tQ \Psi + \Psi\Psi=0\,.
\end{equation}
The gauge transformation is also assumed to be identical in form to the bosonic one,
\begin{equation}
\label{oldGaugeTrans}
\delta \Psi =\tQ \Lambda + [\Psi,\Lambda]\,.
\end{equation}

Before we continue to identify an action, from which~(\ref{oldEOM}) can be derived, an important observation is in order.
While for the cohomology problem, i.e., for the free theory, string fields that are restricted to a fixed picture number are all equivalent,
this is no longer the case now:
Consider a string field $\Psi$ with a fixed picture number.
If $pic(\Psi)=0$ the EOM decomposes into two picture components,
\begin{equation}
Q\Psi+\Psi\Psi=0\,,\qquad \eta \Psi=0\,,
\end{equation}
that is, $\Psi\in \cH_S$ and it obeys the expected EOM.
Similarly, if $pic(\Psi)=-1$ the EOM decomposes into 
\begin{equation}
\eta\Psi+\Psi\Psi=0\,,\qquad Q \Psi=0\,,
\end{equation}
that is, $\Psi$ lives in the dual small Hilbert space and obeys the expected EOM.
On the other hand, for any other value of the picture number, we obtain as a component of the picture decomposition,
\begin{equation}
\Psi\Psi=0\,,
\end{equation}
which seems to be too restrictive. This observation is also consistent with the form of the gauge transformation~(\ref{oldGaugeTrans}),
which inevitably adds to string fields with a fixed picture number $pic(\Psi)\neq 0, -1$ new picture components.
This implies that a gauge fixing to a given picture number is generically inconsistent and while fixing the
(NS sector of the) democratic theory to $pic(\Psi)= 0$ or $pic(\Psi)= -1$ leads to familiar consistent theories,
other gauge fixings must still include string field components at infinitely many different picture numbers.

While the EOM~(\ref{oldEOM}) and gauge transformation~(\ref{oldGaugeTrans}) take the same form as in the bosonic case,
the action from which they are derived cannot be of the same form.
This stems from the fact that we work in the large Hilbert space, in which the total ghost number of an object that gives
a non-trivial CFT expectation value is 2 instead of 3. Also, its parity must be even instead of odd.
Hence, it was assumed that the action contains an odd ghost number $-1$ insertion, $\cO$.
In light of the special role of the mid-point in formulations that include the star product it was assumed that $\cO$ is inserted at the mid-point.
This choice is consistent with the form of the EOM and the gauge transformation
if the following identity holds,
\begin{equation}
[\tQ,\cO]=0\,.
\end{equation}
Furthermore, since it is inserted at the mid-point, $\cO$ must be a zero weight primary.
Also, $\cO$ must carry zero momentum.
All this requirements fix the form of $\cO$ almost uniquely\footnote{The operators $\cO_p$ are defined
only up to the addition of terms of the form $Q\eta \Upsilon_{p+1}$. However, one can impose a universality
condition that fixes these terms as well.} to be,
\begin{equation}
\label{democraticO}
\cO = \sum_{p=-\infty}^\infty \cO_p = \cO_- - \cO_+\,,
\end{equation}
where $\cO_\pm$ are given in~(\ref{Opm})\,.
All in all, the action takes the form,
\begin{equation}
\label{oldAction}
S=\frac{1}{2}\vev{\cO\Big(\Psi\tQ\Psi+\frac{2}{3} \Psi\Psi\Psi\Big)},
\end{equation}
where the expectation values are evaluated in the large Hilbert space and are defined to be non-zero
only for a total picture number $-1$ and total ghost number $2$,
\begin{equation}
\vev{A,B}\equiv\sum_p\vev{A_p, B_{-1-p}}_{BPZ}\,,
\end{equation}
and $\vev{\cdot,\cdot}_{BPZ}$ stands for the usual BPZ inner product.

The EOM that follows from~(\ref{oldAction}) is,
\begin{equation}
\label{oldEOMwithO}
\cO\Big(\tQ \Psi + \Psi\Psi\Big)=0\,.
\end{equation}
This appears not to be quite the same as~(\ref{oldEOM}).
However, it is also assumed that the space of string fields is implicitly defined in a way that
prevents string fields from having mid-point support. In fact, this is a requirement
that should be imposed in any string field theory based on the star product, since otherwise the star
product itself could lead to unacceptable singularities.
With this assumption, the EOM~(\ref{oldEOMwithO}) is equivalent to the EOM~(\ref{oldEOM}).
The gauge transformation derived from the action~(\ref{oldAction})
is exactly~(\ref{oldGaugeTrans}) without any further assumptions.
The fact that the gauge symmetry does not include any additional operators is what distinguishes
this democratic theory from theories such as~\cite{Witten:1986qs,Preitschopf:1989fc,Arefeva:1989cp}
that are rendered inconsistent by their gauge transformations.
Contrary to this, the mid-point insertions in the action do not harm the consistency of the theory
in light of the assumption that string fields can have no mid-point support. The action insertion is
not iterated and hence causes no harm.

The form of the gauge transformation~(\ref{oldGaugeTrans}) implies that in the interacting case
an arbitrary range of picture numbers for the string fields is generically unacceptable.
Assume that this is not the case and let $\Psi_m$ be the component of
$\Psi$ whose picture is minimal and $\Psi_M$ be the component of
$\Psi$ whose picture is maximal. If we assume that $\Lambda$ is allowed to carry pictures in the range
$p< 0$ the second term in~(\ref{oldGaugeTrans}) would generically lead to picture number below $p=m$.
Similarly, $p>0$ components in $\Lambda$ would lead to an increase in picture number above $p=M$.
Thus, a finite range of picture numbers would leave us only with $\Lambda_0$. But then, in order for
the argument regarding the cohomology to hold, we would have to assume that only $\Psi_{-1}$ or $\Psi_0$ are non-zero.
We can assume that the $\Lambda_0$ component is necessary, since without it at least some of the $\Psi$
components would not have a non-linear part in their gauge transformation.
All that leads to the following possibilities\footnote{One could debate whether it is indeed necessary to
impose that all the components of the string field have a non-linear part in their gauge transformation.
We prefer to consider here only the simplest options.
Also, note that we assume here the form of the EOM and gauge symmetry of the old democratic theory.
When one considers insertions that are not restricted to the mid-point the interactions and gauge symmetries
might include picture changing operators and there might be many other legitimate picture ranges for both
$\Psi$ and $\Lambda$. Indeed, below we consider a range of pictures that does not strictly fall under the
current classification.
}:
\begin{enumerate}
\item $\Psi$ and $\Lambda$ take values over all picture numbers.
\item $pic(\Psi)\geq -1$ and $pic(\Lambda)\geq 0$.
\item $pic(\Psi)\geq -1$ and $pic(\Lambda)\geq -1$ with the additional constraint $\eta \Lambda_{-1}=0$.
			At the non-interacting level (that is, when we consider only the first term in~(\ref{oldGaugeTrans})),
			we can write in this case $\Lambda_{-1}=\eta \Upsilon_0$. Then,
			$\tQ \eta \Upsilon_0=Q\eta\Upsilon_0=\eta(-Q\Upsilon_0)$. Hence, we can trade this $\Lambda_{-1}$ by
			$\Lambda_0\equiv -Q\Upsilon_0$, which brings us back to the previous option.
\item $pic(\Psi)\leq 0$ and $pic(\Lambda)\leq 0$. This option is dual under the $Q\Leftrightarrow\eta$
			symmetry to the second option.
\item $pic(\Psi)\leq 0$ and $pic(\Lambda)\leq 1$ with the additional constraint $Q \Lambda_1=0$.
			Similar remark to that of case 3 apply here.
\item $pic(\Psi)= 0,1$ and $pic(\Lambda)=0$.
			This is a two-state solution to the problem
			of constructing superstring field theory. The EOM can be decomposed to,
			\begin{equation}
			\label{twoStateEOM}
			\eta \Psi_{-1}+\Psi_{-1}\Psi_{-1}=0\,,\qquad Q \Psi_0+\Psi_0\Psi_0=0\,,\qquad \eta \Psi_0+Q\Psi_{-1}+[\Psi_0,\Psi_{-1}]=0\,.
			\end{equation}
			and the gauge symmetry is,
			\begin{equation}
			\label{twoStateGauge}
			\delta \Psi_0=Q\Lambda + [\Psi_0,\Lambda]\,,\qquad \delta \Psi_{-1}=\eta\Lambda + [\Psi_{-1},\Lambda]\,.
			\end{equation}
\end{enumerate}
Only the first of these options has been considered in the old democratic theory~\cite{Kroyter:2009rn,Kroyter:2010rk}.
Options 2-5 are not really democratic since only half of the picture numbers can now influence the physics.
We call such options ``pseudo-democratic''\footnote{In fact, many established democracies were only pseudo-democratic
for various reasons for a significant part of their history. Even an established democracy such as Switzerland
was only pseudo-democratic until 1971. Other countries remain pseudo-democratic to this day.}.
In the following we do not consider the last possibility and concentrate on generalizing only the democratic and pseudo-democratic theories.

As already mentioned, one should specify a definition for the space of string fields, since there are many
ways to incorporate a space that includes all picture numbers. Note, that the space $\cH_{com}$~(\ref{Hcom}), for example, is not adequate
for defining a democratic theory, since while the infinitesimal form of the gauge transformation~(\ref{oldGaugeTrans})
leaves a compactly supported state in $\cH_{com}$, the finite form of a gauge transformation does not.

Finally, an important advantage of the democratic theory is the straightforward incorporation of the Ramond sector.
Take $\Psi=\Psi_{NS}+\Psi_R$ where the NS string field $\Psi_{NS}$ includes all possible integer picture numbers
and the Ramond string field $\Psi_R$ includes all possible half-integer picture numbers with the
same action~(\ref{oldAction}). The resulting theory produces the expected equations of motion and gauge symmetries for both sectors
while naturally unifying them.

%==============================================
\subsection{Criticism of the democratic theory}
%==============================================
\label{sec:crit}

Since its introduction the democratic theory was criticized on several grounds.
We mention here some of the arguments against the theory and attempt to address those within
the new approaches developed in the rest of this paper:
\begin{itemize}
\item As already mentioned, an explicit definition of a space of string fields which includes all picture numbers
			but retains the standard cohomology problem was lacking. Actually, similar issues afflict all string field theories,
			in light of the lack of a canonical positive definite norm in these spaces.
\item The democratic theory includes negative weight states. Such states exist in any case for non-zero momentum,
			but now they exist even for zero momentum. As	the picture number $p$
			goes to $\pm\infty$ more such states appear with the conformal weight being unbounded from below even for zero momentum.
\item PCOs do not appear in the gauge transformation, but they do appear in the action
			(within the $\cO_p$ operators). While this does not lead immediately to inconsistencies, it poses a challenge for the perturbative approach to the theory. Perturbation theory requires gauge fixing and the evaluation of
			a propagator. In the current case gauge fixing should also fix the picture number
			(or	remove in any other way the picture number degeneracy) of equivalent states. Finding a consistent way of doing
			so seems not to be trivial. In particular,
			simple fixings to a given value of the picture number lead to expressions in which the (mid-point) PCOs appear
			also in the EOM and the residual gauge transformation, that is, these partial gauge fixings do not appear to be consistent
			ways for handling the gauge symmetry related to picture changing. While one could claim that these gauge fixings are singular limits
			of more regular prescriptions, these regular prescriptions are lacking.
\end{itemize}

%======================
\section{Preliminaries}
%======================
\label{sec:preliminaries}

Much of the criticism on the democratic theory concentrated on the mid-point insertions used to define it.
Hence, we would like to attempt the construction of theories without such insertions.
However, there are some tools that we need before we can attempt to define such theories:
In the old democratic formulation described above, the equivalence of the EOM~(\ref{oldEOMwithO}) and~(\ref{oldEOM})
stems from the assumption that degrees of freedom inserted at the mid-point decouple from degrees of freedom
describing string fields. Thus, the string fields that multiply each mid-point insertion component should
separately vanish. This would no longer be the case for a regular theory
and we would have to find a way to eliminate the insertions
needed for defining the action. But even before that, we have to define these insertions.

In order to overcome the subtleties related to the mid-point we want to define a democratic theory with smeared
insertions, as in the $A_\infty$ formulations. There are many ways to replace the $\cO_p$ mid-point insertions.
One possibility would be to write an integral of the $\cO_p(z)$ operators. However, since these operators are
essentially given by~(\ref{OxiX}), we could replace the mid-point $\cO_p$ by a product of contour integrals
of the form $\xi X^{p-1}$. To that end we define
\begin{subequations}
\label{XYxi}
\begin{align}
\xi & \equiv\frac{1}{2\pi i}\oint \frac{\xi(z)dz}{z}\,,\\
X & \equiv\frac{1}{2\pi i}\oint \frac{X(z)dz}{z}\,.
\end{align}
This definition implies,
\begin{equation}
X=[Q,\xi]\,,
\end{equation}
and therefore $X$ and $\xi$ commute.
The operators $X$ and $\xi$ are just the zero modes of $X(z)$ and $\xi(z)$ respectively. The contour integral
simply instructs us to disregard any singularity from the OPE of $X(z)$ and an operator insertion at the origin.
In addition to positive powers of $X$ we would need negative powers of $X$. This would be the case even when we choose
to construct a pseudo-democratic theory, whose action can be defined only with positive powers of $X$.
The most natural candidate for $X^{-1}$ is $Y$ and we define it similarly,
\begin{equation}
Y \equiv\frac{1}{2\pi i}\oint \frac{Y(z)dz}{z}\,.
\end{equation}
\end{subequations}
With this definition, the regular OPE of $X(z)Y(w)$ implies that
\begin{equation}
\label{XYcom}
[X,Y]=0\,.
\end{equation}
The choice~\eqref{XYxi} for $X$, $\xi$, and $Y$ is not unique and one could generalize the prescription by introducing a function,
subject to some constraints, in the definition of these objects, as was done in~\cite{Erler:2013xta}.
We refrain from that in this work and utilize the zero modes defined above, similarly to our choice of $\eta$,
which is the zero mode of the $\eta(z)$ field\footnote{There is of course an important difference, which we must keep in mind:
$\eta(z)$ is a conformal primary of weight 1 and hence $\eta$ is a derivation, like $Q$. This is not the case with $X$, $\xi$, and $Y$.}.
However, it is important to remember that this is an option for further generalizations.
We also note that $X$, $\xi$, and $Y$ are BPZ even.

The prescription presented here for the PCOs can be expressed in term of the usual normal ordering $::$ as
\begin{subequations}
\label{Order1}
\begin{eqnarray}
X^p V&\equiv& \underset{p}{\underbrace{:X:X\ldots :X}}V\underset{p}{\underbrace{:\ldots :}}\,,\\
Y^p V&\equiv& \underset{p}{\underbrace{:Y:Y\ldots :Y}}V\underset{p}{\underbrace{:\ldots :}}\,.
\end{eqnarray}
\end{subequations}
This is different from another one that can be often seen in the literature,
\begin{subequations}
\label{Order2}
\begin{eqnarray}
\tilde X^p V&\equiv& :\underset{p-1}{\underbrace{:X:X\ldots :X}}X\underset{p-1}{\underbrace{:\ldots :}}V:\,,\\
\tilde Y^p V&\equiv& : \underset{p-1}{\underbrace{:Y:Y\ldots :Y}}Y\underset{p-1}{\underbrace{:\ldots :}}V:\,.
\end{eqnarray}
\end{subequations}
The on-shell associativity of the normal ordered product implies that the two definitions are equivalent inside
the cohomologies of $Q$ and $\eta$ \cite{Lian:1992mn,Lian:1995wa}.
The main advantage of the choice \eqref{XYxi} is that it implies the simple product properties
\begin{subequations}
\begin{eqnarray}
X^p X^q&=&X^q X^p = X^{p+q}\,,\\
Y^p Y^q&=& Y^q Y^p=Y^{p+q}\,.
\end{eqnarray}
\end{subequations}

However, as already mentioned, $X$ has a kernel.  Hence, $Y$ cannot be expected to behave as $X^{-1}$ over arbitrary string fields.
To overcome this difficulty one can consider the restriction of the space of string fields to a subspace $\cH_{res}$ over which the equation
\begin{equation}
\label{XYinv}
X^{-1}= Y\,,
\end{equation}
does hold. It is not immediately clear that this is a well defined statement, nor that this space is large enough to be considered
an off shell extension of the space of NS vertex operators. We examine this issue in section~\ref{sec:XY}.
We do note, already at this stage, that a projection to a space for which
\begin{equation}
\label{XY1}
XY=1\,,
\end{equation}
implies that there is an isomorphism between
different pictures at the level of string fields (i.e., of vertex operators) and not just at the level of the cohomology.
Hence, with such a projection zero-momentum states with negative conformal weight are completely absent,
since they are absent at the $-1$ picture and conformal weights are invariant under the action of $X$ and $Y$.
Moreover, arbitrarily fixing the values of norms of (the finitely many) fixed conformal weight zero-momentum operators at any given
picture number, induces norms for string fields at arbitrary picture number.
This enables a proper definition of string field spaces that are not compactly supported, but decay as a function of picture number,
in a given manner.

As stated above, given the EOM we would have to be able to cancel the insertion factor in order to obtain
the expected form of the EOM. As we shall later see the insertion includes a factor of,
\begin{equation}
\label{Dinst}
\delta(1-X)\equiv \sum_{p=-\infty}^\infty X^p\,,
\end{equation}
in the case of a democratic theory or,
\begin{equation}
\label{PDinst}
\frac{1}{1-X}\equiv \sum_{p=0}^\infty X^p\,,
\end{equation}
in the case of a pseudo-democratic theory. The notation in~(\ref{PDinst}) is self-explanatory. Let us understand the notation~(\ref{Dinst}).
If $X$ was a number we could have tried to evaluate the series by splitting it,
\begin{equation}
\sum_{p=-\infty}^\infty X^p=\sum_{p=0}^\infty X^p + X^{-1}\sum_{p=0}^{\infty} \big(X^{-1}\big)^p=\frac{1}{1-X}+\frac{1}{X}\frac{1}{1-X^{-1}}\,.
\end{equation}
It seems that the result vanishes, but one has to remember that the intersection of the range of convergence of the two series is empty.
Moreover, at $X=1$ it is clear that the expressions do not make sense.
Thus, the result, if meaningful at all, must be proportional to a delta function (or some other generalized distribution concentrated
at $X=1$). One could attempt to extract this part of the expression by separating the expressions into a principal part and a delta function,
but in light of the empty intersection of the two series and the fact that $X$ is not really a real variables that can be integrated
we do not attempt this approach. Instead, we note that the following holds,
\begin{equation}
(1-X) \sum_{p=-\infty}^\infty X^p=0\,,
\end{equation}
by considering the component at each picture number separately.
Thus we conclude that, up to normalization that we do not attempt to
fix\footnote{The standard normalization is anyway related to the integration measure, while $X$ is not a real variable.},~(\ref{Dinst})
makes sense.

The $\delta(1-X)$ factor cannot be inverted. Thus, a democratic theory with such a factor that is inserted not at the mid-point would
lead to EOM that is not equivalent to the expected EOM, at least not in a straightforward way.
The $(1-X)^{-1}$ factor on the other hand can be easily inverted by multiplying it by $1-X$. Thus, the pseudo-democratic construction
seems to be simpler to handle. We attempt to construct the free part of such a theory in section~\ref{sec:free}
and the lowest interacting order in two different ways in sections~\ref{sec:cubic} and~\ref{sec:partialFixed}.

The $\delta(1-X)$ factor suggests yet another possible direction towards the construction of a theory.
This insertion formally has as its kernel all states that are eigenstates of $X$ with unity eigenvalue\footnote{Not to be confused
with the perfectly regular eigenstates with eigenvalue 1 of the operator $XY$ to be considered in the next section.}.
Such states must be nonzero for all integer picture numbers. Moreover, the ``norm'' of a particular picture component is the same for all
picture numbers, i.e., they do not decay as a function of picture number for $p\rightarrow \pm \infty$.
Hence, such states should not be allowed in the space of string fields in the standard approach.
However, the fact that components at neighbouring picture numbers are related by $X$ suggests that they carry exactly the same information.
This statement becomes precise if we again consider states only in the subspace for which $XY=1$, since then $X$ defines
a homomorphism already for states and a unity eigenvalue of $X$ implies also a unity eigenvalue of $Y$.
Since in this case all pictures can contribute to the physics, but they all must
contribute exactly in the same way, we refer to such theories as ``people's democratic'' theories
and to the space of such states as $\cH_{PD}$.
This situation opens other possibilities for defining string vertices. The simplest of the possibilities it to
define vertices that give the same values for states that are related by powers of $X$ or $Y$.
These vertices are well defined over $\cH_{PD}$ if we evaluate them with an arbitrary component,
since they are independent of representatives.
We briefly sketch ideas regarding such a construction in section~\ref{sec:peoples}.
There might also be other ways for defining vertices over this space.

%==========================
\section{The operator $XY$}
%==========================
\label{sec:XY}

As mentioned, the condition~(\ref{XYinv}) is not a trivial one, since PCOs have
non trivial kernels. These kernels cannot include physical states, in light of the fact that these operators define
isomorphisms among the cohomologies at different picture numbers~\cite{Horowitz:1988ip}.
However, this fact by itself does not imply~(\ref{XYinv}) even for physical states.

To understand this point we want to be able to diagonalize the operator $XY$ and to examine its spectrum.
However, it is not clear even whether $XY$ is diagonalizable. Since $XY$ does not change
picture number, ghost number, conformal weight, and momentum, one can analyse it separately in subspaces with fixed
values of these quantum numbers. These are finite dimensional spaces.
The BPZ conjugation defines a bilinear product that couples such spaces in pairs,
with the usual restrictions on the sums of the quantum numbers.
We can strip the $\delta(p_1-p_2)$ factor from the momentum dependence, and remain with finite expressions for
the BPZ bilinear product. With the reality condition enforced the bilinear product defines also a non-positive definite
(but also non-degenerate) inner product~\cite{Witten:1986qs} on each pair of spaces with fixed quantum numbers.
$X$ and $Y$ commute and are both self-adjoint under the BPZ conjugation.
Hence, had the inner product been positive definite it would have implied that $XY$ is diagonalizable.

In any case, over each one of these finite dimensional subspaces $XY$ can at least be put in Jordan form.
Which eigenvalues can this operator obtain?
We know that the spectrum includes 0 and 1. If $XY$ was diagonalizable and these were its only eigenvalues it would have been
a projector\footnote{There are other variants of the picture changing operator that lead to an operator $XY$,
which is a projector~\cite{Erler:2016ybs}. However, in this case $X^2$ is singular and the operators $X$ and $Y$ do not
commute. Hence, this choice is not adequate for a democratic formulation.}.
This is not the case, as can be seen by explicitly evaluating some particular examples.
Consider the state\footnote{We do not distinguish states from the operators producing them under the
state-operator correspondence.} $(b\xi')(0)$. Direct evaluation shows that,
\begin{equation}
XY(b\xi')=X\big(-\xi''\xi'e^{-2\phi}\big)=6b\xi'-2b\xi'=4(b\xi')\,,
\end{equation}
that is, the eigenvalue is $4$.
As another example, consider the state $(c\eta)(0)$. Now,
\begin{equation}
XY(c\eta)=X\Big(\big(e^{-2\phi}\big)'c'c+\frac{1}{2}e^{-2\phi}c''c\Big)=-4c\eta+c\eta=-3(c\eta)\,.
\end{equation}
Thus, $-3$ is also an eigenvalue of $XY$.
Adding momentum to this example we obtain,
\begin{equation}
XY(c\eta e^{ik\cdot x})=X\Big(\Big(\big(e^{-2\phi}\big)'c'c+\frac{1}{2}e^{-2\phi}c''c\Big)e^{ik\cdot x}\Big)
    =\big(-3c\eta-2e^{-\phi}c'c(\psi\cdot k)\big)e^{ik\cdot x}\,.
\end{equation}
We see that with non-zero momentum this state is no longer an eigenstate.
Consider now,
\begin{equation}
XY\Big(-2e^{-\phi}c'c(\psi\cdot k)e^{ik\cdot x}\Big)=X\Big(-c''c'c \xi' e^{-3\phi}(\psi\cdot k)e^{ik\cdot x}\Big)
    =-2e^{-\phi}c'c(\psi\cdot k)e^{ik\cdot x}\,.
\end{equation}
Thus, this is an eigenstate with eigenvalue $-2$.
The matrix representation of $XY$ restricted to these two states in the basis they naturally define can be written as,
\begin{equation}
XY|_{2\mbox{ states}}=\left(\begin{array}{cc}-3 &0 \\ 1 &1\end{array}\right),
\end{equation}
which can easily be diagonalized.

We want to note some properties of the eigenvalues of $XY$: If $V$ is an eigenstate with eigenvalue $\lambda$,
so are also $X^n V$ and $Y^n V$ for all $n>0$, in light of the commutativity of $X$ and $Y$, as long as they are non-zero.
States of the form $X^n V$ and $Y^n V$ can equal zero for $V$ an eigenstate of $XY$ only if $\lambda=0$,
since otherwise we obtain, e.g.,
\begin{equation}
Y^n (X^n V)=(XY)^n V=\lambda^n V\neq 0\,.
\end{equation}
Similarly, if $V$ is an eigenstate with eigenvalue $\lambda$ and $QV\neq 0$,
then $QV$ is an eigenstate with eigenvalue $\lambda$, since $X$ and $Y$ commute with $Q$.

Physical states $V$ which are eigenstates of $XY$ must have eigenvalue 1. To show that, let us consider a generic
expectation value involving a physical state $V$, living in the small Hilbert space, $\vev{V(0)...}$.
Here, the dots stand for arbitrary other physical operators, inserted in other points and the expectation value is in the small Hilbert space.
Since all states are on-shell, we can insert $X(z)Y(w)$ for any choice of $z,w$ in such an expression, without changing its value.
Then, since $V$, as well as the other vertices in this expression are physical, we can move $X$ and $Y$ arbitrarily obtaining,
\begin{equation}
\begin{aligned}
\vev{V(0)...} &=\vev{X(z)Y(z+\epsilon)V(0)...}=\vev{X(z)Y(\epsilon)V(0)...}=\vev{X(z)(YV)(0)...}=\\
              &=\vev{(XYV)(0)...}=\lambda \vev{V(0)...}\,.
\end{aligned}
\end{equation}
Here, the third equality comes from the fact that, as $\epsilon$ approaches $0$, the difference between
the two expressions contains only the contribution of singular terms, which must be exact and hence do not contribute to the result,
which must be independent of the location of $Y$. In the next equality the same logic was applied to $X$.
Now either $\vev{V(0)...}=0$ for all possible insertions, which is possible only if $V$ is exact, or $\lambda=1$.
In the former case we have an exact state, which can have an arbitrary eigenvalue, while in the latter, we have a physical state,
whose eigenvalue must equal unity.
Note, that this expression does not by itself imply that $V$ is an eigenstate, since the inner product in the space of physical states
is non-degenerate only modulo $Q$-exact terms. Hence, in general one could expect to have,
\begin{equation}
XY V=V+Q\Lambda\,,
\end{equation}
for some $\Lambda$.
If this is the case it might happen that $V$ is a generalized eigenstate belonging to a Jordan chain starting with some exact state,
$Q\Lambda$, whose eigenvalue is 1.

The situation of a physical state belonging to the Jordan chain of an exact state seems to us somewhat unnatural.
Furthermore, if this turns out to be the case, it might be possible to modify the choice of $X$ and $Y$, by choosing
a specific function in the generalization described in~\cite{Erler:2013xta}, such that $V$ would be an eigenstate.
While this issue certainly deserves further study, we now {\it assume} that there exists a choice under which all physical states
are indeed genuine eigenvalues of $XY$. With this assumption, despite the fact that $XY$ itself is not a projector, we can define
a projector $\cP$, to the space of genuine eigenstates with $\lambda=1$, which under our assumption includes all the physical states.
We then declare that the string field lives in the restricted space defined by,
\begin{equation}
\label{Hres}
\cH_{res}\equiv \cP \cH_L\,,
\end{equation}
or in terms of the states,
\begin{equation}
\label{VHres}
V\in \cH_{res} \quad \Longleftrightarrow \quad V=\cP V \in \cH_L \,.
\end{equation}

Defining vertices over the linear space $\cH_{res}$ is simple: Take any string vertex and replace everywhere
\begin{equation}
\label{PsiPPsi}
\Psi \rightarrow \cP \Psi\,.
\end{equation}
In this way we can also interpret the string vertices as defined over the whole space $\cH_L$\footnote{Presumably
restricted to a given ghost number and with some proper definition of behaviour as a function of picture number.}.
Then, all states that do no obey~(\ref{VHres}) decouple, i.e., they lead to zero result. This implies that states that
are killed by the projector can be interpreted as pure gauge states.
Since we use~(\ref{PsiPPsi}) for all the string vertices, this new gauge symmetry remains exact at the interacting level as well.
In particular, we see that the restriction~(\ref{Hres}) does not miss any gauge symmetries. Moreover, in light of the discussion above,
the states that are added to the list of pure gauge states are not physical and hence the cohomology problem is intact.

The primary string vertex that should be defined is one that corresponds to the symplectic form over the space $\cH_{res}$.
In particular, it must be non-degenerate. Consider the standard BPZ inner product in $\cH_L$.
One can define bases with respect to which the BPZ inner product takes the form\footnote{This is not necessarily a real basis,
since there might be a sign difference between the BPZ and the Hermitian conjugation.},
\begin{equation}
\vev{\tilde V_i,V_j}=\delta_{ij}\,,
\end{equation}
where $\tilde V_i$ and $V_j$ live in conjugate spaces, in the sense of the quantum numbers mentioned above.
If we restrict the $V_j\in \cH_{res}$ to a basis for the subspace of $\cH_{res}$ with the given quantum numbers, we obtain,
\begin{equation}
\delta_{ij}=\vev{\tilde V_i,V_j}=\vev{\tilde V_i,\cP V_j}=\vev{\cP \tilde V_i,V_j},
\end{equation}
where in the last equality we used the fact that $\cP$ is a function of the BPZ even $XY$.
We infer that the $\hat V_i\equiv\cP \tilde V_i$ form a basis for the space dual to the one of the $V_j$
and are also in $\cH_{res}$.
Thus, the BPZ inner product is non-degenerate also when restricted to $\cH_{res}$.
From this we infer that if we write an action, with all string fields obeying the projection condition,
a variation of the action that can be brought to the form,
\begin{equation}
\vev{\delta \Psi,F(\Psi)}=0\,,
\end{equation}
for some functional $F$ of the string field, is equivalent (in light of the non-degeneracy of the inner product) to the EOM,
\begin{equation}
\cP F(\Psi)=0\,.
\end{equation}
Also, (unless we work in $\cH_L$ with the new gauge symmetry related to the projector, mentioned above)
any gauge transformation constructed for the field $\Psi$ must obey,
\begin{equation}
\delta_{\mbox{gauge}}\Psi=\cP \delta_{\mbox{gauge}}\Psi\,.
\end{equation}

%========================
\section{The free action}
%========================
\label{sec:free}

The free action is expected to take the form,
\begin{equation}
\label{S0action}
S_2\equiv \frac{1}{2}\vev{\cO\Psi,\tQ\Psi},
\end{equation}
where $\Psi$ is a string field.
Following the previous discussion we have several options for defining this action.
In particular,  one can consider a fully democratic theory or a pseudo-democratic one. While the former choice is more symmetric,
the latter has the advantage of having an insertion $\cO$, whose dependence on $X$ is invertible.
Also, one has to decide whether to impose $\Psi\in\cH_{res}$ or not.
Again, the more natural choice seems to be not to impose it, but imposing it does not modify the cohomology problem,
while enabling us to use~(\ref{XY1}).
With these considerations in mind we choose here to impose the projection to $\cH_{res},$\footnote{Since all the operations
in this section keep states in $\cH_{res}$ we do not write explicitly the projection $\cP$ to this space.
In particular note that this projection commutes with the projection to a given picture number or a range of picture numbers used below.}
and consider the pseudo-democratic case, that is, we define,
\begin{equation}
\Psi=\sum_{p=-\infty}^{-1}\Psi_p\,,
\end{equation}
where $p$ is the picture number and the ghost number is restricted to $g=1$.
Similarly, $\cO$ is defined as,
\begin{equation}
\cO\equiv \sum_{p=1}^\infty\cO_p\,,\qquad \cO_p\equiv \xi X^{p-1}\,,
\end{equation}
where $\xi$ and $X$ are defined in~(\ref{XYxi}).
The operator $\cO$ is odd, BPZ even and obeys,
\begin{equation}
\label{contHom}
[\tQ,\cO]=-1\,,
\end{equation}
that is, it is (the negative of) a contracting homotopy operator\footnote{It is an integrated version of~(\ref{Op}).} for the derivation $\tQ$.

The variation of the action gives,
\begin{equation}
\begin{aligned}
\label{varS0}
\delta S_2 &= \frac{1}{2}\Big(\vev{\cO\delta\Psi,\tQ\Psi}+\vev{\cO\Psi,\tQ\delta\Psi}\Big)
						=\frac{1}{2}\Big(-\vev{\delta\Psi,\cO\tQ\Psi}-\vev{\tQ\cO\Psi,\delta\Psi}\Big)=\\
					 &=\frac{1}{2}\Big(-\vev{\delta\Psi,\cO\tQ\Psi}+\vev{\cO\tQ\Psi,\delta\Psi}+\vev{\Psi,\delta\Psi}\Big)=-\vev{\delta\Psi,\cO\tQ\Psi}.
\end{aligned}
\end{equation}
Here, we used in the second equality the BPZ property of $\cO$ and the fact that both $\cO$ and $\Psi$ are Grassmann odd,
as well as integrated by parts the derivation $\tQ$, then, we used the commutation relation~(\ref{contHom}).
Finally, we used the fact that the picture number of the $\Psi\delta\Psi$ is bounded from above by $-2$, so this term vanishes.

Since $\mathrm{pic\left(\delta\Psi\right)\leq-1}$, the variation implies that the components of the expression that
multiplies this variation vanish at picture number zero and higher. Thus, we obtain the following EOM,
\begin{equation}
\cP_{p\geq 0} \big(\cO \tQ\Psi\big)=0\,,
\label{eom0}
\end{equation}
where $\cP_{p\geq 0}$ is a projector operator to the given range of picture numbers. This can be written explicitly as,
\begin{equation}
\sum_{q=-\infty}^{-2}\sum_{p=-q}^{\infty}\cO_{p}\left(Q\Psi_{q}-\eta\Psi_{q+1}\right)+\sum_{p=1}^{\infty}\cO_{p}Q\Psi_{-1}=0\,.
\end{equation}
Decomposing to different picture numbers we obtain,
\begin{equation}
\forall p\geq 0\quad \cO_{p+1}Q\Psi_{-1}+\sum_{q=-\infty}^{-2}\cO_{p-q}\left(Q\Psi_{q}-\eta\Psi_{q+1}\right)=0\,.
\end{equation}
This looks like an infinite set of equations. However, in $\cH_{res}$~(\ref{XY1}) implies
that these equations are in fact dependent. Multiplying the equation for $p>0$ by $Y$ gives the $p-1$ equation.
Thus, the set of infinitely many equations is equivalent to the $p=0$ equation,
\begin{equation}
\label{singleEOM}
\xi\Big(Q\Psi_{-1}+\sum_{q=-\infty}^{-2} X^{-q-1}\left(Q\Psi_{q}-\eta\Psi_{q+1}\right)\Big)=0\,.
\end{equation}

The EOM has the following gauge invariance,
\begin{equation}
\label{gauge0}
\delta\Psi=\tQ\Lambda+\xi\hat\Lambda\,,\qquad pic\big(\Lambda\big)\leq -1\,,\qquad pic\big(\hat\Lambda\big)\leq -2\,.
\end{equation}
The $\Lambda$ gauge invariance follows from the equation of motion~(\ref{eom0}) upon using~(\ref{Qeta}).
It can also be inferred from the variation of the action~(\ref{varS0}) using ``integration by parts'' and ignoring again the $\vev{\Psi,\delta\Psi}$
term whose picture is bounded from above by $-2$. As for the $\hat \Lambda$ transformation, using
\begin{equation}
[\tQ,\xi]=X-1\,,
\end{equation}
and
\begin{equation}
\label{XO}
X \cO_p=\cO_{p+1}\,,
\end{equation}
the gauge transformation of the equation of motion~(\ref{eom0}) is,
\begin{equation}
\cP_{p\geq 0} \big(\cO \big(X-1-\xi\tQ\big)\Lambda\big)=-\cP_{p\geq 0} \big(\xi\Lambda\big)=0\,.
\end{equation}
Here, we used in the first equality the fact that $\cO$ is proportional to $\xi$ which leads to,
\begin{equation}
\label{Oxi}
\cO \xi=0\,,
\end{equation}
as well as the identity,
\begin{equation}
\cO(X-1)=-\xi\,.
\end{equation}
Then, in the second equality, the projection gives zero, since $pic(\xi\hat\Lambda)\leq -1$.
Verifying that this transformation is indeed a gauge symmetry using the variation of the action is even simpler,
since all that is needed is a single integration by parts in~(\ref{varS0}) and the identity~(\ref{Oxi}).

Note, that we could have also allowed in~(\ref{gauge0}) a gauge transformations with
$pic(\Lambda)=0$ provided that $Q\Lambda$ vanishes. These are indeed gauge transformations. However, for such a gauge transformation
the triviality of $Q$ in the large Hilbert space implies that there exists a state $\Omega$ obeying $\Lambda=Q\Omega$.
We can therefore write,
\begin{equation}
\tQ\Lambda=(Q-\eta)Q\Omega=-\eta Q\Omega=Q(\eta\Omega)=\tQ(\eta\Omega)\,,
\end{equation}
where $pic(\eta\Omega)=-1$. Hence, the incorporation of these gauge fields does not change the set of allowed gauge transformations,
but only the set of gauge fields. We can use the $\hat\Lambda$ gauge transformation in order to fix $\Psi\in \cH_S$.
Now $\eta\Psi=0$ and the $\xi$ in front of the equation of motion~(\ref{singleEOM}) can
be removed by acting with $\eta$ on the equation. Hence, we are left with,
\begin{equation}
Q\Psi=0\,,\qquad \Psi\in \cH_S\,,\qquad pic(\Psi)\leq -1\,.
\end{equation}
The coefficients of $\Psi$ at picture numbers less than $-1$ can be eliminated using the $\Lambda$ gauge symmetry,
since for $p\leq 2$ we can set $\Lambda=\xi \Psi_p$, which leads to
\begin{equation}
\delta \Psi=(X-1)\Psi_p - \xi(Q-\eta)\Psi_p\,.
\end{equation}
The last term takes us away from the fixed $\hat \Lambda$ gauge. Hence, this transformation should be accompanied
by a $\hat \Lambda$ transformation that would exactly remove this term and we are left with a transformation
that raises the picture of the $\Psi_p$ component by one.
Hence, the given cohomology problem is equivalent to a cohomology problem for which the EOM is\footnote{The proposed
sequence of gauge transformations leads to a state in $\cH_{-1}$ which is defined by a series, with the series coefficients
being defined by the various gauge transformations. This series can be decomposed with respect to a standard basis of the Hilbert space
to give a set of number series for the coefficients of the basis states. A possible definition for the space of string fields
in this pseudo-democratic theory is the space of all string fields that lead in this way only to coefficient series that are
absolutely convergent.},
\begin{equation}
Q\Psi=0\,,\qquad \Psi\in \cH_S\cap \cH_{res}\,,\qquad pic(\Psi)= -1\,,
\end{equation}
and the remaining gauge symmetry is,
\begin{equation}
\delta \Psi=Q\Lambda\,.
\end{equation}
Now, $pic(\Lambda)=-1$ and $\Lambda\in \cH_S\cap \cH_{res}$, since otherwise we would also have to consider the $\eta\Lambda$
contribution, which would lead to a $pic=-2$ component that would take us away from the partial gauge fixing.
All in all, we see that a partial gauge fixing leads exactly to the standard cohomology problem.
Hence, the current formulation, defined by the EOM~(\ref{eom0}) and the gauge
transformation~(\ref{gauge0}) over a space restricted by~(\ref{XY1}), is equivalent to it.
Thus, we can take the free action~(\ref{S0action}) together with the linear constraint~(\ref{XY1}) as
a starting point for defining an interacting superstring field theory.

%=========================
\section{The cubic vertex}
%=========================
\label{sec:cubic}

So far we constructed only the free theory.
What one would like to do now is to extend it to an interacting theory.
To that end we would like to add higher order vertices to the action, such that the following would hold:
\begin{enumerate}
\item The action would lead to EOM of the form\footnote{We attempt a perturbative addition of higher order terms.
			In what follows, we only attempt to examine the first not trivial order, namely, the cubic vertex.
			In principle it might happen that this is all that is needed. It might also happen that higher order terms would also be needed.
			This might lead to an interesting mathematical structure, e.g., an $A_\infty$ algebra, but we do not examine this issue here.},
			\begin{equation}
			\label{cubicEOM}
			Z\big( \tQ\Psi+M_2(\Psi,\Psi)+\ldots \big)=0\,,
			\end{equation}
			where $M_2$ is some extension of the star product, the dots stand for higher order terms (we leave $g_s$ implicit, but the expansion is clear),
			and $Z$ is some linear operator, which could in principle be unity, that does not alter the expected form of the EOM at least for some large
			enough sector of string fields, e.g., small Hilbert space states $\Psi$ with $pic(\Psi)=-1$.
\item When restricted to physical states in the small Hilbert space the three vertex gives the expected values for scattering of string states,
			regardless of picture numbers of the states and even if $\Psi$ includes several picture components.
\item The linearized gauge symmetries of the previous section should either be shown to hold without modification when the higher
			order terms are added, or be properly modified with their own higher order terms.
\end{enumerate}
Let us recall that the above mentioned linearized gauge symmetries consist of the following:
\begin{itemize}
\item The usual gauge symmetry of the small Hilbert space, at any fixed picture number. These gauge symmetries are generated by $\tQ$ which
			equals $Q$ when acting on small Hilbert space states.
\item The extension of the previous gauge symmetry to the large Hilbert space. This includes the symmetries that are responsible for picture changing,
			generated by $X-1$ and $Y-1$, for on-shell small Hilbert space states.
\item The gauge symmetry that is generated by $\xi$, which trivializes the ``upper component'' of the large Hilbert space.
			In addition to leaving only the small Hilbert space as the physical one, this symmetry, together with the previous one, also enables
               the picture changing of off-shell small Hilbert space states.
\item The restriction to $\cH_{res}$ can be replaced by a gauge symmetry that trivializes the other component of the string field.
			As was shown above, this gauge symmetry appears when we add projection operators to $\cH_{res}$ on all entries of the string vertices.
			We continue to assume that these projectors appear everywhere and keep them implicit. This implies that this gauge symmetry is not being
			modified at the non-linear level. Furthermore, we can assume that we fixed it and simply work in the space $\cH_{res}$.
\end{itemize}

In fact, a trivial construction obeying all the mentioned requirements exists:
At the quadratic level we showed that one can use the gauge symmetries in order to transform the whole
string field to the small Hilbert space at any particular picture number.
We can declare that the gauge symmetries used to that end are unchanged at the non-linear level and for each
string field $\Psi$ construct a gauge equivalent state $\psi$,
which lives in the small Hilbert space at any particular picture number, e.g., at $pic(\psi)=-1$. We can then extend the residual gauge symmetry,
as well as the action, to incorporate the $A$-infinity structure of EKS~\cite{Erler:2013xta}.
Such a theory, while not completely identical to that of~\cite{Erler:2013xta}, in light of the restriction to $\cH_{res}$, would have all the
needed properties.
Furthermore, such a construction can work also for a fully democratic theory, not only for the pseudo-democratic case advocated above.
Nonetheless, it essentially consists of what might be seen as an artificial extension by gauge degrees of freedom
of a known construction and it is not fundamentally democratic.
We would like to examine the possibility of a more universal (with respect to picture number) extension of the linearized theory.

In order to obtain~(\ref{cubicEOM}) we want to generalize the variation of the free theory~(\ref{varS0})
to the following variation when the cubic term is added,
\begin{equation}
\label{var23}
\delta S_{2;3} = -\vev{\delta\Psi,\cO\big(\tQ\Psi+M_2(\Psi,\Psi)\big)}.
\end{equation}
Note that in this case the gauge symmetry related to $\xi$ remains exact at the cubic order.
This is desirable, since $\xi$ is not a derivation and constructing a non-trivial non-linear generalization thereof might be challenging.
Furthermore, this idea is easily generalized also to higher order vertices.

The suggested form of the variation~(\ref{var23}) fixes the operator $Z$ of~(\ref{cubicEOM}),
\begin{equation}
\label{Z}
Z=\cP_{p\geq -1} \cP_S \frac{1}{1-X}\,,
\end{equation}
where $\cP_S=\eta\xi$ is a projection to the small Hilbert space. Note, that the factor of $(1-X)^{-1}$ brings different picture components
to the same value of picture number, so they will appear in the same picture-projected equation. Different picture components can be present
in~(\ref{cubicEOM}) since $\Psi$ does not have to be concentrated at a single picture number and in general $M_2$ as well is expected to map
string fields with various picture components to a string  field having several picture components. 
If $pic\big(M_2(\Psi,\Psi)\big)\leq -1$ we can simplify the form of $Z$ to
\begin{equation}
\label{Zt}
Z=\cP_{p= -1} \cP_S \frac{1}{1-X}\,,
\end{equation}
since all other picture components of the equation derived using~(\ref{Z}) will be related to the ones derived using~(\ref{Zt})
by extracting extra powers of $X$ from the $(1-X)^{-1}$ factor.
Similarly, if $pic\big(M_2(\Psi,\Psi)\big)\leq q$ for some $q>-1$ we effectively have,
\begin{equation}
\label{Ztt}
Z=\cP_{-1\leq p \leq q} \cP_S \frac{1}{1-X}\,.
\end{equation}
Otherwise, we have to work with~(\ref{Z}). The $-1$ component of the EOM includes is such a case the kinetic term, while at higher
picture number new components, coming from the $M_2$ term, will be added.
These new components essentially introduce some further constraints on the main EOM,
obtained at picture number $-1$.
It is natural to expect that at least for some large enough class of string fields these components would be absent.

Note an important (further) difference between the old democratic theory and the construction we are currently attempting.
Below~(\ref{oldEOM}) we explained that in the interacting old democratic theory there is no symmetry among different picture numbers,
in the sense that non-trivial solutions of the EOM that are concentrated at a single picture number can exist only for $pic(\Psi)=0,-1$.
This is not the case now. The factor of $(1-X)^{-1}$ in~(\ref{Zt}) brings contributions at all picture number to the same equation.
Hence, at least superficially they are equivalent. This suggests that the gauge symmetry related to picture changing is either
not changed at the interacting level, or is changed in a less dramatic way than in the old democratic theory.
We come back to this issue at the end of this section and in the next one.
For now, we want to stress that the difference stems not from the fact that we consider here a pseudo-democratic theory, as opposed to
the fully democratic old theory. In fact, as we already mentioned, one could construct a pseudo-democratic theory from the old theory
by imposing a gauge choice that sets to zero all non-negative picture numbers. In such a theory picture $-1$ would still be different
from all the other pictures for the reasons stated above. The point is rather that in the old theory the location of the insertion
is at the mid-point, which leads, according to our assumptions, to its decoupling from the string fields, leading to an EOM,
with different components at different picture numbers.
Again, as mentioned below~(\ref{XYcom}), one could attempt to generalize our choice of $X$ and $Y$ in a manner similar to
that of~\cite{Erler:2013xta}, by introducing a dependence on some function. Then, in a certain limit one would obtain the construction
presented here, while in another (presumably singular) limit one would obtain a mid-point based formalism, with the mentioned properties
of the old democratic theory.

Let us now attempt to identify an action from which the variation~(\ref{var23}) follows.
To that end, we write the quadratic action as,
\begin{equation}
S_2=\frac{1}{2}\omega(\tQ\Psi,\Psi)\,,
\end{equation}
where $\omega$ is the would be symplectic form -- it is degenerate over gauge orbits, but reduces to a proper symplectic form after gauge fixing
the $\xi$ symmetry and considering the small Hilbert space. It takes the form (the inner product is still in the large Hilbert space),
\begin{equation}
\label{omegaDef}
\omega(A,B)\equiv \vev{\cO A,B}.
\end{equation}
Note, that the BPZ properties of $\cO$ imply that $\omega$ is symmetric\footnote{We do not attempt at this stage to identify an $A_\infty$ structure.
Hence, we use Grassmann parity rather than the degree in the exponents of $(-1)$. Similarly, below, we do not modify the star product with a minus sign.
These conventions could be easily modified if needed.},
\begin{equation}
\label{omegaSym}
\omega(A,B)=(-1)^{AB}\omega(B,A)\,.
\end{equation}

For on-shell small Hilbert space states the expected scattering amplitudes would be obtained if we
define\footnote{The lowest $X$ power does not contribute in this case, so one can replace
$\cO$ by $\frac{\xi X}{1-X}$, but one could as well write the expression in its current form.},
\begin{equation}
\label{M2_m2}
S_{3,on-shell}\simeq\frac{1}{3}\omega\big(\Psi,m_2(\Psi,\Psi)\big)\,,
\end{equation}
where we now write Witten's star product explicitly as $m_2$. For on-shell states it is not important where one inserts the $\cO$ operator.
Hence, when restricted to these states~(\ref{var23}) is obtained from this action.
However, this is no longer the case for general states -- while $m_2$ is cyclic with respect to $\omega$ for on-shell small Hilbert space
states, this is not the case in general. Hence, we look for $M_2\simeq m_2$ that would be cyclic in the general case, i.e.,
we want it to satisfy,
\begin{equation}
\label{M2def}
\omega\big(A,M_2(B,C)\big)=(-1)^{A(B+C)}\omega\big(B,M_2(C,A)\big)\,,
\end{equation}
and define the cubic term as,
\begin{equation}
\label{S3}
S_3=\frac{1}{3}\omega\big(\Psi,M_2(\Psi,\Psi)\big)\,.
\end{equation}
In order to show that the property~(\ref{M2def}) implies that the variation of the action takes the desired form~(\ref{var23}),
we first note that,
\begin{equation}
\label{M2cyc}
\omega\big(A,M_2(B,C)\big)=(-1)^{C(A+B)}\omega\big(C,M_2(A,B)\big)=\omega\big(M_2(A,B),C\big)\,.
\end{equation}
Here, we used~(\ref{M2def}) twice in the first equality, and used~(\ref{omegaSym}) in the second equality.
Using~(\ref{M2def}) and~(\ref{M2cyc}) it follows immediately that,
\begin{equation}
\begin{aligned}
\delta S_3 & =\frac{1}{3}\Big(\omega\big(\delta\Psi,M_2(\Psi,\Psi)\big)+\omega\big(\Psi,M_2(\delta\Psi,\Psi)\big)
		+\omega\big(\Psi,M_2(\Psi,\delta\Psi)\big)\Big)=\\
		& =\omega\big(\delta\Psi,M_2(\Psi,\Psi)\big)\,,
\end{aligned}
\end{equation}
as desired.

For defining $M_2$ that obeys~(\ref{M2cyc}) we have to be able ``to move'' the location of the $\cO$ insertion
in the definition of $\omega$.
There are several operators that can be used to that end\footnote{This is in contrast to the situation we would have
faced with a fully democratic theory, since the operator~(\ref{democraticO}) has no inverse.}.
We define the first term in the definition of the vertex in the following way\footnote{We will shortly explain why
more terms are needed and identify these terms.},
\begin{equation}
M_{2,1}(B,C)=\frac{1}{3}R\Big(\cO m_2(B,C)+m_2(\cO B,C)+(-1)^B m_2(B,\cO C)\Big)\,,
\end{equation}
where $R$ is a BPZ-odd Grassmann-odd operator that can be used to remove, using BPZ conjugation, the $\cO$ insertion on the
definition of $\omega$,
\begin{equation}
\label{RO}
[R,\cO]=1\,.
\end{equation}
Moreover, $R$ should eliminate physical small Hilbert space states, in order not to modify the scattering amplitudes.
There are several options for choosing $R$\footnote{These operators are BPZ odd when we include the minus sign stemming from
the ``integration by parts'' (change of the contour direction) in the definition of their BPZ conjugation.
One could consider also BPZ even operators for defining $R$, but we identify no natural candidates in this case.}, e.g.,
\begin{subequations}
\begin{align}
& R_1 = -\tQ\,,\\
& R_2 = (Y-1)Q\,,\\
& R_3 = (1-X)\eta\,.
\end{align}
\end{subequations}
In fact, we can choose,
\begin{equation}
\label{generalR}
R=\alpha R_1+\beta R_2 +(1-\alpha-\beta)R_3\,,
\end{equation}
for arbitrary values of $\alpha,\beta$. Note also that $R$ is nilpotent regardless of the choice of $\alpha,\beta$,
\begin{equation}
R^2=0\,.
\end{equation}
The choice,
\begin{equation}
\label{RtQ}
R=R_1 = -\tQ\,,
\end{equation}
has the added advantage of $R$ being a derivation. However, for now we continue with the general form of $R$.

Of course, for small Hilbert space physical states we can move the location of the insertion even with $m_2$ as a product.
Our goal is to be able to do that for general states. Thus, we would have to add more terms in the definition of the vertex.
In order to identify the next term in the definition of the vertex, we plug the first one into the action and obtain,
\begin{align}
\nonumber
& \omega\big(A,M_{2,1}(B,C)\big) = \frac{1}{3}(-1)^A\vev{R\cO A,\cO m_2(B,C)+m_2(\cO B,C)+(-1)^B m_2(B,\cO C)}=\\
\label{M21res}
				 & \qquad = \frac{1}{3}(-1)^A\vev{A,\cO m_2(B,C)+m_2(\cO B,C)+(-1)^B m_2(B,\cO C)}\\
\nonumber
				 & \qquad \quad -\frac{1}{3}(-1)^A\vev{\cO RA,m_2(\cO B,C)+(-1)^B m_2(B,\cO C)},
\end{align}
where in the first equality we used~(\ref{omegaDef}) and integrated $R$ by parts, and in the second equality
we used~(\ref{RO}), as well as the nilpotency of $\cO$ in order to eliminate one term from the last line.
The term in the second line is manifestly cyclic and leads to the sough for interaction for on-shell small
Hilbert space states. However, the term in the last line, which vanishes for on-shell small Hilbert space states, is not cyclic.

To remedy this we note that the problematic terms of the last line of~(\ref{M21res}) include one insertion of $\cO$ and one insertion of $\cO R$.
There are four additional ways to distribute these insertions (without acting at the same location). We have to add to $M_2$
a second term that would produce these insertions (with the proper signs) inside the cubic action~(\ref{S3}).
Again, we can use $R$ in order to eliminate the $\cO$ that comes from the definition of $\omega$, where needed.
Thus, we propose to add to the vertex the following expression,
\begin{align}
& M_{2,2}(B,C)=\\
\nonumber
& \quad -\frac{1}{3}\Big(m_2(\cO R B,C) + m_2(B,\cO R C) + (-1)^B R m_2(\cO R B,\cO C) + R m_2(\cO B, \cO R C) \Big).
\end{align}
This part gives,
\begin{align}
\omega\big(A,M_{2,2}(B,C)\big) = - &\frac{1}{3} \Big(\!
\vev{\cO A,m_2(\cO R B,C)} + \vev{\cO A,m_2(B,\cO R C)} \\
\nonumber
+& (-1)^{A+B} \vev{A,m_2(\cO R B,\cO C)} + (-1)^A\vev{A,m_2(\cO B,\cO R C)} \\
-& (-1)^{A+B} \vev{\cO R A,m_2(\cO R B,\cO C)} -(-1)^A \vev{\cO R A,m_2(\cO B,\cO R C)}
\!\Big).
\nonumber
\end{align}
The first two lines of this expression combine with~(\ref{M21res}) to a manifestly cyclically symmetric expression.
The last line is not cyclic, but this can be remedied by adding a third piece to $M_2$,
\begin{equation}
M_{2,3}(B,C) = \frac{1}{3} m_2(\cO R B,\cO R C)\,,
\end{equation}
which leads to,
\begin{equation}
\omega\big(A,M_{2,3}(B,C)\big) = \frac{1}{3}\vev{\cO A, m_2(\cO R B,\cO R C)}\,.
\end{equation}

We managed to construct a fully symmetric vertex. This vertex was not uniquely defined, due to the ambiguity
in the definition of $R$~(\ref{generalR}). Furthermore, we also have the freedom of adding another piece to the vertex,
which would have no effect on the resulting action (and is therefore of no importance at this stage),
\begin{equation}
M_{2,4}(B,C)=\cO \hat M_2(B,C)\,,
\end{equation}
with $\hat M_2$ being completely arbitrary (except for, e.g., its ghost number), in light of the nilpotency of $\cO$.
Again, this would not have been possible in a fully democratic theory, since there, $\cO^2$ is not well defined.
Similarly, the expressions obtained now in the action, which include several insertions of $\cO$ would not have been
well defined off-shell in a fully democratic theory, but are well defined in the current case, since at each picture number $p$
the coefficient of $X^p$ (with the $X$'s being arbitrarily inserted) is well defined and finite.

We obtained an action from which~(\ref{cubicEOM}) is obtain as the EOM.
This is a non-linear extension of~(\ref{eom0}), which we proved to be equivalent to the standard cohomology problem.
Since~(\ref{cubicEOM}) originates from an action that gives the expected results for on-shell scattering amplitudes,
we expect it to be consistent.
Note that the equation depends implicitly on the choice of $R$.
Presumably, different choices of $R$ are related to each other by some complicated field redefinition accompanied by a gauge transformation.

Of course, the non-linear gauge transformation itself depends on the choice of $R$ and we still have to identify it.
It seems that the simplest possibility for choosing $R$ is~(\ref{RtQ}), in light of the derivation property
and since $\tQ$ is the operator that appears in the linear term.
Thus, we assume the choice~(\ref{RtQ}) below.
Recall that the gauge symmetry related to $\xi$ remain exact at the cubic order.
Hence, we only need to identify the non-linear extension of the gauge symmetry related to $\tQ$.
The equation we have to solve is,
\begin{equation}
\label{secondOrderGauge}
\omega\big(\tQ\Lambda,M_2(\Psi,\Psi)\big)+\omega(\delta_2 \Psi,\tQ \Psi)=0\,.
\end{equation}
This expression superficially resembles the expressions obtained in the bosonic and in the cubic NS theories.
Hence, one could think that the solution should be similar, i.e.,
\begin{equation}
\label{WrongExpect}
\delta_2\Psi\stackrel{?}{\sim} M_2(\Psi,\Lambda)-M_2(\Lambda,\Psi)\,.
\end{equation}
However, this cannot be the case for several reasons. First, $\omega$ includes the $\cO$ factor that does not
commute with $\tQ$. Moreover, $\tQ$ (or any other choice of $R$ we considered) is not a derivation of $M_2$.
Finally and most importantly, the picture number of $\delta_2 \Psi$ should, like that of $\Psi$, be bounded from above by $-1$,
while the picture numbers resulting from $M_2$ are unbounded from above.
Hence, gauge transformations of the form~(\ref{WrongExpect}) might have been considered in a fully democratic theory,
but not at the case at hand.

In order to identify the non-linear term in the gauge transformation, one first has to evaluate,
\begin{equation}
\label{gauge2ndP2}
\vev{\tQ \Lambda,\cO M_2(\Psi,\Psi)} = \vev{\Lambda,M_2(\Psi,\Psi)+\cO \tQ M_2(\Psi,\Psi)}.
\end{equation}
Using~(\ref{RtQ}) we immediately obtain,
\begin{equation}
\tQ M_{2,1}(B,C)=0\,.
\end{equation}
Similarly, the last two terms of $M_{2,2}$ drop away and we are left with,
\begin{equation}
\begin{aligned}
\tQ M_{2,2}(B,C)=- \frac{1}{3}\Big(& m_2(\tQ B,C) + (-1)^B m_2(B,\tQ C)\\
	-& (-1)^B m_2(\cO \tQ B, \tQ C) - m_2(\tQ B,\cO \tQ C)\Big).
\end{aligned}
\end{equation}
Finally,
\begin{equation}
\tQ M_{2,3}(B,C) = - \frac{1}{3} \Big(m_2(\tQ  B,\cO \tQ C) + (-1)^B m_2(\cO \tQ  B,\tQ C)\Big).
\end{equation}
All in all we obtain\footnote{This is an interesting expression. It appears as if $\tQ$ turns $M_2$ into $m_2$
and then acts on it as a derivation, up to a multiplication by a constant. We are not sure of the importance of this observation.},
\begin{equation}
\tQ M_2(B,C)=- \frac{1}{3}\Big(m_2(\tQ B,C) + (-1)^B m_2(B,\tQ C)\Big).
\end{equation}
In particular,
\begin{equation}
\label{OQM2contrib}
\vev{\Lambda, \cO \tQ M_2(\Psi,\Psi)}=- \frac{1}{3}\vev{\Lambda, \cO m_2(\tQ \Psi,\Psi) - \cO m_2(\Psi,\tQ \Psi)}.
\end{equation}

Next, we examine the contributions to the term $\vev{\Lambda,M_2(\Psi,\Psi)}$, appearing in~(\ref{gauge2ndP2}),
again term by term,
\begin{equation}
\begin{aligned}
\label{M21contrib}
M_{2,1}(\Psi,\Psi) =  - & \frac{1}{3}\tQ\Big(\cO m_2(\Psi,\Psi)+m_2(\cO \Psi,\Psi) - m_2(\Psi,\cO \Psi)\Big)=\\
										& \frac{1}{3}\Big(3m_2(\Psi,\Psi)+\cO m_2(\tQ\Psi,\Psi)- \cO m_2(\Psi,\tQ\Psi)+\\
										& + m_2(\cO \tQ\Psi,\Psi)+ m_2(\Psi,\cO \tQ\Psi)-m_2(\cO \Psi,\tQ\Psi)+ m_2(\tQ\Psi,\cO \Psi)\Big).
\end{aligned}
\end{equation}
Note that the first term in the second line drops out when plugged into~(\ref{secondOrderGauge}) due to its picture number range,
while the next two terms exactly cancel out the contribution from~(\ref{OQM2contrib}).
Then,
\begin{align}
\nonumber
M_{2,2}&(\Psi,\Psi) =
\frac{1}{3}\Big(2 m_2(\cO \tQ \Psi,\Psi) +2 m_2(\Psi,\cO \tQ \Psi) + m_2(\cO \Psi, \tQ \Psi) -  m_2(\tQ \Psi,\cO \Psi)+\\
&+ 2 m_2(\cO \tQ \Psi,\cO \tQ \Psi) \Big).
\label{M22contrib}
\end{align}
Here, the first two terms in the first line combine nicely with the first two terms in the last line of~(\ref{M21contrib}),
while the other two terms exactly cancel out the last two terms of~(\ref{M21contrib}).
Finally, we have,
\begin{equation}
\begin{aligned}
M_{2,3}&(\Psi,\Psi) = \frac{1}{3} m_2(\cO \tQ \Psi,\cO \tQ \Psi)\,,
\end{aligned}
\end{equation}
which combines nicely with the term in the second line of~(\ref{M22contrib}).

We now see that~(\ref{secondOrderGauge}) implies,
\begin{equation}
\label{delta2}
\vev{\delta_2\Psi,\cO\tQ\Psi} + \vev{\Lambda,m_2(\cO \tQ\Psi,\Psi)+ m_2(\Psi,\cO \tQ\Psi) + m_2(\cO \tQ \Psi,\cO \tQ \Psi)}=0\,.
\end{equation}
Canceling the first two terms of the second expression is straightforward and we obtain the standard form for the gauge transformation,
\begin{equation}
\delta_{2,1}\Psi=m_2(\Psi,\Lambda) - m_2(\Lambda,\Psi)\,.
\end{equation}
Canceling the last term is much harder. It might seem that we can choose,
\begin{equation}
\label{WrongExpect2}
\delta_{2,2}\Psi\stackrel{?}{=} \alpha m_2(\cO \tQ \Psi, \Lambda) - (1-\alpha) m_2(\Lambda,\cO \tQ \Psi)\,,
\end{equation}
for an arbitrary $\alpha$. In fact, this cannot be the case, since $\Psi$ (again) would carry a wrong
range of picture numbers.
The fact that $\delta_2\Psi$ enters~(\ref{delta2}), in which it is contracted by $\cO$ that includes a factor of $(1-X)^{-1}$
might seem to suggest that we could lower the picture of~(\ref{WrongExpect2}) and this would be compensated by some power of
$X$ already present in the expression. This is also not quite correct, since lowering the picture of~(\ref{WrongExpect2})
would also introduce new couplings that are not present, to states with high enough pictures. In fact, it seems that if
one attempts to naively lower the picture of all the components of~(\ref{WrongExpect2}) in order to keep the picture number in
the legitimate range, the result might not converge. Thus, it is not clear to us what is the form of the second part of the gauge
transformation at the cubic order. Presumably an expression exists that is non universal in its treatment of picture numbers
that utilizes picture changing in some more sophisticated way.

%=============================
\section{Partial gauge fixing}
%=============================
\label{sec:partialFixed}

Since the $\hat \Lambda$ gauge symmetry~(\ref{gauge0}) remains exact in the construction, one can consider the possibility of
fixing it and constructing an interacting theory for the fixed case.
The most natural way to fix this symmetry is to restrict $\Psi$ to the small Hilbert space.
To obtain the form of the residual linear gauge symmetry we split $\Lambda$ into its two components
and rewrite~(\ref{gauge0}) as,
\begin{equation}
\label{gauge_part_fixed}
\delta\Psi=\tQ(\Lambda+\xi \tilde\Lambda)+\xi\hat\Lambda\,,
\end{equation}
with $\Lambda,\tilde\Lambda,\hat\Lambda\in \cH_S$.
Setting,
\begin{equation}
\hat\Lambda = Q \tilde\Lambda\,,
\end{equation}
we obtain a variation that leaves the string field in the small Hilbert space.
The residual gauge symmetry takes the form,
\begin{equation}
\label{gauge_part_fixed1}
\delta\Psi=Q\Lambda+(X-1) \tilde\Lambda\,.
\end{equation}
We recognize the standard gauge symmetry generated by $\Lambda$ as well as a picture changing symmetry
generated by $\tilde\Lambda$. In particular, choosing $\tilde\Lambda=\Psi$ raises the picture number of
the given string field and choosing $\tilde\Lambda=-Y\Psi$ lowers it.
In order to remain in the appropriate picture range for the string fields one has to impose
\begin{equation}
pic\big(\Lambda\big)\leq -1\,,\qquad pic\big(\tilde\Lambda\big)\leq -2\,,
\end{equation}
in accord with the picture range defined in~(\ref{gauge0}).

The free action takes now the form,
\begin{equation}
\label{S0actionF}
S_2 = \frac{1}{2}\vev{\frac{1}{1-X}\Psi,Q\Psi} \equiv \frac{1}{2} \omega(\Psi,Q\Psi)\,,
\end{equation}
where now the expectation value is taken in the small Hilbert space and a variation of the action leads to
\begin{equation}
Q\Psi=0\,.
\end{equation}
The factor of $(1-X)^{-1}$ in the action exactly guarantees that the EOM be obeyed by any picture components in the range $pic(\Psi)\leq -1$.
Again, we would like to search for a cubic (and eventually higher order) extension of the action that would lead to EOM of the form,
\begin{equation}
Z\big(Q\Psi+M_2(\Psi,\Psi)\big)=0\,,
\end{equation}
for a product $M_2$ generalizing the star product and some proper operator $Z$.
Again, a trivial solution exists: using the picture changing gauge symmetry in order to squeeze the string
field to its $-1$ picture component and then using an analogue of the EKS action~\cite{Erler:2013xta}.
We would like to find a more universal treatment of the string field.
However, for $p<-1$ the factor of $(1-X)^{-1}$ becomes non-trivial and the location of its insertion
starts to play a role.

A simple non-trivial choice for $M_2$ exists, with a uniform distribution of the insertion
among the three locations,
\begin{equation}
M_2(B,C)=\frac{1-X}{3}\left((1-X)^{-1} m_2(B,C) +m_2((1-X)^{-1} B,C) + m_2(B, (1-X)^{-1} C)\right).
\end{equation}
Note that for this $M_2$ the kinetic operator $Q$ is a derivation.
The cubic part of the action is again of the form of~(\ref{S3}),
\begin{equation}
S_3=\frac{1}{3}\omega\big(\Psi,M_2(\Psi,\Psi)\big) = \frac{1}{3}\vev{\frac{1}{1-X}\Psi,m_2(\Psi,\Psi)},
\end{equation}
and we recognize that now,
\begin{equation}
Z=\cP_{-1\leq p} \frac{1}{1-X}\,.
\end{equation}
Similar remarks to the ones mentioned in the previous section hold here as well.

Inspecting the gauge transformation in the current case we obtain,
\begin{equation}
\omega\big(\delta_2\Psi+M_2(\Lambda,\Psi)-M_2(\Psi,\Lambda)\big)=0\,.
\end{equation}
Again, this seems to suggest an expression for the gauge transformation,
\begin{equation}
\delta_2\Psi\stackrel{?}{=} M_2(\Psi,\Lambda) - M_2(\Lambda,\Psi)\,,
\end{equation}
which is again unacceptable in light of the allowed range of picture numbers.
In a sense it even seems that the problem becomes now more severe,
presumably since we attempted to fix the $\xi$ gauge symmetry in an inconsistent way.

%===================================================
\section{A note on a ``people's democratic theory''}
%===================================================
\label{sec:peoples}

At the end of section~\ref{sec:preliminaries} we mentioned the option of a ``people's democratic theory'',
which is constructed over string fields $\Psi\in\cH_{PD}$, where this space of string fields is defined by,
\begin{equation}
\label{XPsiPsi}
X\Psi=\Psi\,.
\end{equation}
Note that this also implies $Y\Psi=\Psi$ in light of~(\ref{XYcom}) and~(\ref{XY1}).
Such string fields carry exactly the same information in all their picture components.
The fact that we have in such a string field infinitely many copies of the same physical object suggests that
a regular construction of a kinetic term would lead to divergences. However, in light of~(\ref{XPsiPsi})
we have a natural way to define a kinetic term. Again, as demonstrated in the previous subsections, we can attempt
the construction of a theory either in the small or in the large Hilbert space.
We present here only a sketch of the construction in the small Hilbert space, for simplicity.

The free action can be written as,
\begin{equation}
\label{pdAction2}
S_2=\frac{1}{2}\vev{\delta(1-X)\Psi_p,Q\Psi_q},
\end{equation}
where now the expectation value is evaluate in the small Hilbert space and the picture numbers $p$ and $q$
are arbitrary.
In light of the defining property~(\ref{XPsiPsi}) and the presence of the delta factor in the action
it is immediate that this definition is independent of the representatives $p$ and $q$ and thus it makes sense.
What we did essentially is to factor out the infinities that would have multiplies this action had we
attempted to write $\Psi$ instead of $\Psi_p$ and $\Psi_q$.

A variation of the action~(\ref{pdAction2}) leads immediately to\footnote{One should not be confused by the different roles of $\delta$
as a formal picture-number delta-function and as the variation of $\Psi$.},
\begin{equation}
\delta S_2=\frac{1}{2}\Big(\vev{\delta(1-X)\delta\Psi_p,Q\Psi_q} + \vev{\delta(1-X)\delta\Psi_q,Q\Psi_p}\Big)\,,
\end{equation}
which we can rewrite using the fact that this expression is independent of $p$ and $q$ as
\begin{equation}
\delta S_2=\vev{\delta(1-X)\delta\Psi_p,Q\Psi_q} = \vev{X^{-2-p-q}\delta\Psi_p,Q\Psi_q}=
\vev{\delta\Psi_{-1},Q\Psi_{-1}},
\end{equation}
or in any other desired picture.
This variation leads immediately to
\begin{equation}
Q\Psi_{-1}=0\,,
\end{equation}
which together with~(\ref{XPsiPsi}) implies
\begin{equation}
Q\Psi=0\,,
\end{equation}
as we should have. Similarly, the linearized gauge symmetry is the expected one,
\begin{equation}
\delta\Psi=Q\Lambda\,,
\end{equation}
where $\Lambda$ also obeys
\begin{equation}
X\Lambda=\Lambda\,.
\end{equation}

The construction of the kinetic term cannot be directly applied to cubic and higher order terms. Now, there are several possibilities
for inserting the $X$ powers appearing in the $\delta(1-X)$ factor, which are not equivalent for off-shell states.
Moreover, different distribution of the picture number among the three string field insertions could also lead to different results,
i.e., now there is a dependence on the representatives.
One possible way for defining a canonical interaction term would be to average over all possibilities. However, the picture number
in a discrete variable, and as is well known there is no uniform distribution over the integers.
Nonetheless, one could hope to define the action as a limit. The kinetic term~(\ref{pdAction2}) could be written as,
\begin{equation}
S_2=\frac{1}{2}\lim_{N\rightarrow\infty}\frac{1}{(2N+1)^2}\sum_{p,q=-N}^N\vev{X^{-2-p-q}\Psi_p,Q\Psi_q}.
\end{equation}
In fact, this expression equals~(\ref{pdAction2}) even before taking the limit, in light of the independence on representatives.
However, this suggests defining,
\begin{equation}
S_3\equiv \frac{1}{3}\lim_{N\rightarrow\infty}\frac{1}{(2N+1)^3}\sum_{p,q,r=-N}^N\vev{X^{-2-p-q-r}\Psi_p,m_2(\Psi_q,\Psi_r)}.
\end{equation}
Alternatively, one could define a smoother limit\footnote{See~\cite{Fuchs:2006gs} for an example of the necessity of using a smooth limit
in another context of regularizing string field theory.},
\begin{equation}
S_3\equiv \frac{1}{3}\lim_{\alpha\rightarrow 1^-}\frac{(1-\alpha)^3}{8}
\sum_{p,q,r=-\infty}^\infty\alpha^{|p|+|q|+|r|}\vev{X^{-2-p-q-r}\Psi_p,m_2(\Psi_q,\Psi_r)}.
\end{equation}
One could further average over the possible locations of the $X$ insertions, i.e., an $X^2$ factor could be inserted at
a single point, as we wrote above, but it could also be split to two $X$ factors that are inserted at two different points.
One could even have a $Y$ factor in one point and factors of $X$ and $X^2$ in the two other points.
Various averaging of such distributions could be considered.

We defer questions regarding the existence of the limits defining the action, proper distributions of powers of $X$,
the EOM and its gauge transformations, higher order terms, and practical issues such as calculating within such theories to future work.

%====================
\section{Conclusions}
%====================
\label{sec:conclusions}

In this work we examined directions towards the construction of a regular open democratic string field theory.
As a first step we examined possible PCOs that could be used for defining such a theory,
we defined a projection that enables the extension of picture changing to a large enough
class of off-shell states and we examined the properties of the operator defining it.
We identified several possible starting points that could be used for the construction of a theory and concentrated
on the pseudo-democratic case in the NS sector.
We identified the free action both for the unfixed and partially fixed gauge case
and extended it to the first interacting level. However, the identification of the gauge symmetry at this order
proved to be more challenging and the expressions we managed to obtain are incomplete.
Also, the $M_2$ that we obtained leads to terms with arbitrarily high picture number, which seems somewhat unnatural.
We also commented, in the previous section, on another possible direction towards a construction of a theory over
the space $\cH_{PD}$.

There is certainly much more to be done. An obvious future direction would be to reexamine the problems we faced
with the definition of the gauge symmetry. Whereas a possible resolution might emerge from a modification of the PCOs used
or a more sophisticated definition of the product $M_2$, in particular by modifying the choice of the operator $R$,
it might be the case that a more radical change is in order.
As we noticed, the situation with the unfixed theory is somewhat better than that of the partially gauge fixed one.
Presumably, some further gauge freedom or a different treatment of the existing gauge symmetry might save the day.
In our construction we assumed that the gauge symmetry related to $\xi$ does not change at the non-linear level.
This is certainly not a natural assumption. While it would certainly be much more challenging to construct a theory
in which a gauge symmetry that is based on $\xi$, which is not a derivation, attains corrections, this might be necessary.
Another linearity assumption we made is related to the projection to $\cH_{res}$. Again, a more involved
algebraic structure might be in order at the interacting level.

An even bolder treatment of the $\xi$ gauge symmetry can be thought of by noticing an analogy with the starting
point for constructing the Berkovits theory. There, at the linearized level, one has two (anti-)commuting gauge symmetries,
induced by $Q$ and $\eta$. This structure was then extended to the non-linear level.
Now, instead of $Q$ we have $\tilde Q$ and instead of $\eta$ we have $\xi$. Could one generalize the construction
to this case? Again, a most salient obstruction would be the fact that unlike the other mentioned operators, $\xi$
is not a derivation. Also, it is not a primary and it carries the wrong parity.
Furthermore, questions regarding picture number and other restrictions of the space of string fields emerge.
Presumably some coefficient fields would have to carry opposite Grassmann parity
to compensate for the wrong parity of $\xi$. The fact that $\xi$ is not a derivation might lead to a non-associative
algebra. Nonetheless, this is a tempting road to consider. Such line of research might also lead to a better understanding
of the symmetry structure of superstring theory.

While the directions mentioned above are certainly interesting and important they seems to be formidably challenging.
There are also other avenues to pursue. One such important direction is the further study of the $XY$-based projection
and more generally of the properties of the $XY$ operator. Recall that we assumed that it is diagonalizable but did not
provide a proof. There might be subtleties with this assumption, as we described in section~\ref{sec:XY} and if so
there might be various remedies for these subtleties. Moreover one could reexamine the necessity of using the projector
and the possibility to define regular (pseudo-)democratic theories without relying on a projector.
It should be noted in light of one of the examples given in section~\ref{sec:XY}, that the projector might project
some states only at a particular value of the momentum. This seems, again, quite unnatural and should be
further examined.

Another interesting research question is the construction of a two-state theory based on~(\ref{twoStateEOM}) and~(\ref{twoStateGauge}).
It might be the case that a complete construction of such a theory is possible. While such a theory is certainly not too
democratic, it might manifest a symmetry between the $0$ and $-1$ pictures, which are special in some sense, mentioned above.
Moreover, while we are very far from the inclusion of the Ramond sector in the current construction, its inclusion,
with picture number $-\frac{1}{2}$, in such a two-state theory, would probably be relatively straightforward.
Again, this might manifest some underlying symmetry of the superstring.

A symmetry that we see throughout our construction is the symmetry related to interchanging the roles of $X$ and $Y$,
together with some other required changes, e.g., changing the range of picture numbers. One can obtain in such a way
a dual pseudo-democratic theory with an inverted range of picture numbers.
Nonetheless, it is a curious observation that despite this apparent symmetry it is much harder to obtain a fully
democratic formulation, that would further manifest the symmetry between $X$ and $Y$.
Presumably this might have something to do with the asymmetric role that $X$ and $Y$ play in the perturbative
construction of the world-sheet theory of the superstring~\cite{Witten:2012bh}.
We leave this question as well for future work.

%==========================
\section*{Acknowledgements}
%==========================

We would like to thank Harold Erbin, Ted Erler, Udi Fuchs, Carlo Maccaferri, and Ashoke Sen for discussion.
The research presented was supported by the Israel Science Foundation (ISF), grant No. 244/17.
Completion of the project took place while the research of S.G was supported by a BIRD-2021 project (PRD-2021)
and by the PRIN Project n. 2022ABPBEY, ``Understanding quantum field theory through its deformations''. 

%\newpage
\bibliography{bib}

\providecommand{\href}[2]{#2}\begingroup\raggedright\begin{thebibliography}{10}

\bibitem{Witten:1985cc}
E.~Witten, \emph{Noncommutative geometry and string field theory},
  \href{https://doi.org/10.1016/0550-3213(86)90155-0}{\emph{Nucl. Phys. B}
  {\bfseries 268} (1986) 253}.

\bibitem{Witten:1986qs}
E.~Witten, \emph{Interacting field theory of open superstrings},
  \href{https://doi.org/10.1016/0550-3213(86)90298-1}{\emph{Nucl. Phys. B}
  {\bfseries 276} (1986) 291}.

\bibitem{Wendt:1987zh}
C.~Wendt, \emph{Scattering amplitudes and contact interactions in {Witten's}
  superstring field theory},
  \href{https://doi.org/10.1016/0550-3213(89)90118-1}{\emph{Nucl. Phys. B}
  {\bfseries 314} (1989) 209}.

\bibitem{Preitschopf:1989fc}
C.~R. Preitschopf, C.~B. Thorn and S.~A. Yost, \emph{Superstring field theory},
  \href{https://doi.org/10.1016/0550-3213(90)90276-J}{\emph{Nucl. Phys. B}
  {\bfseries 337} (1990) 363}.

\bibitem{Arefeva:1989cp}
I.~Y. Arefeva, P.~B. Medvedev and A.~P. Zubarev, \emph{New representation for
  string field solves the consistency problem for open superstring field
  theory}, \href{https://doi.org/10.1016/0550-3213(90)90189-K}{\emph{Nucl.
  Phys.} {\bfseries B341} (1990) 464}.

\bibitem{Kroyter:2009zi}
M.~Kroyter, \emph{Superstring field theory equivalence: {Ramond sector}},
  \href{https://doi.org/10.1088/1126-6708/2009/10/044}{\emph{JHEP} {\bfseries
  0910} (2009) 044} [\href{https://arxiv.org/abs/0905.1168}{{\ttfamily
  0905.1168}}].

\bibitem{Friedan:1985ge}
D.~Friedan, E.~J. Martinec and S.~H. Shenker, \emph{Conformal invariance,
  supersymmetry and string theory},
  \href{https://doi.org/10.1016/0550-3213(86)90356-1}{\emph{Nucl. Phys. B}
  {\bfseries 271} (1986) 93}.

\bibitem{Berkovits:1995ab}
N.~Berkovits, \emph{Super-{P}oincar\'e invariant superstring field theory},
  \href{https://doi.org/10.1016/0550-3213(95)00259-U}{\emph{Nucl. Phys. B}
  {\bfseries 450} (1995) 90}
  [\href{https://arxiv.org/abs/hep-th/9503099}{{\ttfamily hep-th/9503099}}].

\bibitem{Kroyter:2009rn}
M.~Kroyter, \emph{Superstring field theory in the democratic picture},
  \href{https://doi.org/10.4310/ATMP.2011.v15.n3.a3}{\emph{Adv. Theor. Math.
  Phys.} {\bfseries 15} (2011) 741}
  [\href{https://arxiv.org/abs/0911.2962}{{\ttfamily 0911.2962}}].

\bibitem{Kroyter:2010rk}
M.~Kroyter, \emph{Democratic superstring field theory: {Gauge} fixing},
  \href{https://doi.org/10.1007/JHEP03(2011)081}{\emph{JHEP} {\bfseries 1103}
  (2011) 081} [\href{https://arxiv.org/abs/1010.1662}{{\ttfamily 1010.1662}}].

\bibitem{Kroyter:2011nc}
M.~Kroyter, \emph{Democratic superstring field theory and its gauge fixing},
  \href{https://doi.org/10.1143/PTPS.188.244}{\emph{Prog.Theor.Phys.Suppl.}
  {\bfseries 188} (2011) 244}
  [\href{https://arxiv.org/abs/1101.1314}{{\ttfamily 1101.1314}}].

\bibitem{Erler:2013xta}
T.~Erler, S.~Konopka and I.~Sachs, \emph{Resolving {W}itten's superstring field
  theory}, \href{https://doi.org/10.1007/JHEP04(2014)150}{\emph{JHEP}
  {\bfseries 1404} (2014) 150}
  [\href{https://arxiv.org/abs/1312.2948}{{\ttfamily 1312.2948}}].

\bibitem{Iimori:2013kha}
Y.~Iimori, T.~Noumi, Y.~Okawa and S.~Torii, \emph{From the {Berkovits}
  formulation to the {Witten} formulation in open superstring field theory},
  \href{https://doi.org/10.1007/JHEP03(2014)044}{\emph{JHEP} {\bfseries 03}
  (2014) 044} [\href{https://arxiv.org/abs/1312.1677}{{\ttfamily 1312.1677}}].

\bibitem{Erler:2014eba}
T.~Erler, S.~Konopka and I.~Sachs, \emph{{NS-NS} sector of closed superstring
  field theory}, \href{https://doi.org/10.1007/JHEP08(2014)158}{\emph{JHEP}
  {\bfseries 1408} (2014) 158}
  [\href{https://arxiv.org/abs/1403.0940}{{\ttfamily 1403.0940}}].

\bibitem{Erler:2015lya}
T.~Erler, S.~Konopka and I.~Sachs, \emph{{Ramond} equations of motion in
  superstring field theory},
  \href{https://doi.org/10.1007/JHEP11(2015)199}{\emph{JHEP} {\bfseries 11}
  (2015) 199} [\href{https://arxiv.org/abs/1506.05774}{{\ttfamily
  1506.05774}}].

\bibitem{Erbin:2021smf}
H.~Erbin, \emph{{String Field Theory: A} modern introduction},
  \href{https://doi.org/10.1007/978-3-030-65321-7}{\emph{Springer, vol. 980 of
  Lecture Notes in Physics} (2021) }
  [\href{https://arxiv.org/abs/2301.01686}{{\ttfamily 2301.01686}}].

\bibitem{Maccaferri:2023vns}
C.~Maccaferri, \emph{{String Field Theory}},
  \href{https://arxiv.org/abs/2308.00875}{{\ttfamily 2308.00875}}.

\bibitem{Sen:2024nfd}
A.~Sen and B.~Zwiebach, \emph{{String Field Theory: A Review}},
  \href{https://arxiv.org/abs/2405.19421}{{\ttfamily 2405.19421}}.

\bibitem{Kazama:1985hd}
Y.~Kazama, A.~Neveu, H.~Nicolai and P.~C. West, \emph{{Symmetry Structures of
  Superstring Field Theories}},
  \href{https://doi.org/10.1016/0550-3213(86)90302-0}{\emph{Nucl. Phys. B}
  {\bfseries 276} (1986) 366}.

\bibitem{Kugo:1988mf}
T.~Kugo and H.~Terao, \emph{New gauge symmetries in {Witten's Ramond} string
  field theory},
  \href{https://doi.org/10.1016/0370-2693(88)90640-5}{\emph{Phys. Lett. B}
  {\bfseries 208} (1988) 416}.

\bibitem{Sen:2015hha}
A.~Sen, \emph{Gauge invariant {1PI} effective superstring field theory:
  Inclusion of the {Ramond} sector},
  \href{https://doi.org/10.1007/JHEP08(2015)025}{\emph{JHEP} {\bfseries 08}
  (2015) 025} [\href{https://arxiv.org/abs/1501.00988}{{\ttfamily
  1501.00988}}].

\bibitem{Kunitomo:2015usa}
H.~Kunitomo and Y.~Okawa, \emph{{Complete action for open superstring field
  theory}}, \href{https://doi.org/10.1093/ptep/ptv189}{\emph{PTEP} {\bfseries
  2016} (2016) 023B01} [\href{https://arxiv.org/abs/1508.00366}{{\ttfamily
  1508.00366}}].

\bibitem{Erler:2016ybs}
T.~Erler, Y.~Okawa and T.~Takezaki, \emph{Complete action for open superstring
  field theory with cyclic {$A_\infty$} structure},
  \href{https://doi.org/10.1007/JHEP08(2016)012}{\emph{JHEP} {\bfseries 08}
  (2016) 012} [\href{https://arxiv.org/abs/1602.02582}{{\ttfamily
  1602.02582}}].

\bibitem{Konopka:2016grr}
S.~Konopka and I.~Sachs, \emph{Open superstring field theory on the restricted
  {Hilbert} space}, \href{https://doi.org/10.1007/JHEP04(2016)164}{\emph{JHEP}
  {\bfseries 04} (2016) 164}
  [\href{https://arxiv.org/abs/1602.02583}{{\ttfamily 1602.02583}}].

\bibitem{Kroyter:2012ni}
M.~Kroyter, Y.~Okawa, M.~Schnabl, S.~Torii and B.~Zwiebach, \emph{{Open
  superstring field theory I: gauge fixing, ghost structure, and propagator}},
  \href{https://doi.org/10.1007/JHEP03(2012)030}{\emph{JHEP} {\bfseries 1203}
  (2012) 030} [\href{https://arxiv.org/abs/1201.1761}{{\ttfamily 1201.1761}}].

\bibitem{Berkovits:2012np}
N.~Berkovits, \emph{Constrained {BV} description of string field theory},
  \href{https://doi.org/10.1007/JHEP03(2012)012}{\emph{JHEP} {\bfseries 03}
  (2012) 012} [\href{https://arxiv.org/abs/1201.1769}{{\ttfamily 1201.1769}}].

\bibitem{Berkovits:XXX}
N.~Berkovits, M.~Kroyter, Y.~Okawa, M.~Schnabl, S.~Torii and B.~Zwiebach,
  \emph{{Open superstring field theory II: approaches to the BV master
  action}}, {\emph{unpublished $\!\!\!$} }.

\bibitem{Iimori:2015aea}
Y.~Iimori and S.~Torii, \emph{Relation between the reducibility structures and
  between the master actions in the {Witten} formulation and the {Berkovits}
  formulation of open superstring field theory},
  \href{https://doi.org/10.1007/JHEP10(2015)127}{\emph{JHEP} {\bfseries 10}
  (2015) 127} [\href{https://arxiv.org/abs/1507.08757}{{\ttfamily
  1507.08757}}].

\bibitem{Witten:2012bh}
E.~Witten, \emph{Superstring perturbation theory revisited},
  \href{https://arxiv.org/abs/1209.5461}{{\ttfamily 1209.5461}}.

\bibitem{Sen:2015hia}
A.~Sen and E.~Witten, \emph{Filling the gaps with {PCO's}},
  \href{https://doi.org/10.1007/JHEP09(2015)004}{\emph{JHEP} {\bfseries 09}
  (2015) 004} [\href{https://arxiv.org/abs/1504.00609}{{\ttfamily
  1504.00609}}].

\bibitem{Berkovits:2001us}
N.~Berkovits, \emph{{Relating the RNS and pure spinor formalisms for the
  superstring}},
  \href{https://doi.org/10.1088/1126-6708/2001/08/026}{\emph{JHEP} {\bfseries
  08} (2001) 026} [\href{https://arxiv.org/abs/hep-th/0104247}{{\ttfamily
  hep-th/0104247}}].

\bibitem{Berkovits:2005bt}
N.~Berkovits, \emph{Pure spinor formalism as an {N=2} topological string},
  \href{https://doi.org/10.1088/1126-6708/2005/10/089}{\emph{JHEP} {\bfseries
  10} (2005) 089} [\href{https://arxiv.org/abs/hep-th/0509120}{{\ttfamily
  hep-th/0509120}}].

\bibitem{Lian:1992mn}
B.~H. Lian and G.~J. Zuckerman, \emph{{New perspectives on the BRST algebraic
  structure of string theory}},
  \href{https://doi.org/10.1007/BF02102111}{\emph{Commun. Math. Phys.}
  {\bfseries 154} (1993) 613}
  [\href{https://arxiv.org/abs/hep-th/9211072}{{\ttfamily hep-th/9211072}}].

\bibitem{Lian:1995wa}
B.~H. Lian and G.~J. Zuckerman, \emph{{Algebraic and geometric structures in
  string backgrounds}},  in \emph{{STRINGS 95: Future Perspectives in String
  Theory}}, pp.~323--335, 6, 1995,
  \href{https://arxiv.org/abs/hep-th/9506210}{{\ttfamily hep-th/9506210}}.

\bibitem{Horowitz:1988ip}
G.~T. Horowitz, R.~C. Myers and S.~P. Martin, \emph{{BRST} cohomology of the
  superstring at arbitrary ghost number},
  \href{https://doi.org/10.1016/0370-2693(89)91587-6}{\emph{Phys. Lett.}
  {\bfseries B218} (1989) 309}.

\bibitem{Fuchs:2006gs}
E.~Fuchs and M.~Kroyter, \emph{Universal regularization for string field
  theory}, \href{https://doi.org/10.1088/1126-6708/2007/02/038}{\emph{JHEP}
  {\bfseries 0702} (2007) 038}
  [\href{https://arxiv.org/abs/hep-th/0610298}{{\ttfamily hep-th/0610298}}].

\end{thebibliography}\endgroup

\vfill\eject

\end{document}